\documentclass[preprint,epsf,aps,floats]{revtex4}
\usepackage{amsmath,amssymb,bm,epsfig,psfrag}
\usepackage{mathtools}

\usepackage{color}

\DeclarePairedDelimiter\abs{\lvert}{\rvert}%
\newcommand\+{\dagger}

\newcommand{\be}{\begin{equation}}
\newcommand{\ee}{\end{equation}}
\newcommand{\ber}{\begin{eqnarray}}
\newcommand{\eer}{\end{eqnarray}}

\newcommand\ket[1]{|{#1}\rangle}

\def\Dsl{\,\raise.15ex \hbox{/}\mkern-12.8mu D}

\begin{document}

\title{Singlet Fission in Chiral Carbon Nanotubes: Density Functional Theory Based Computation}
\author{Andrei~Kryjevski,~Deyan~Mihaylov}
\affiliation{Department of Physics,~North Dakota State University,~Fargo, ND~58108,~USA}
\author{Brendan~Gifford,~Dmitri~Kilin,}
\affiliation{Department of Chemistry,~North Dakota State University,~Fargo, ND~58108,~USA}

\begin{abstract}
Singlet fission (SF) process, where a singlet exciton decays into a pair of spin one exciton states which are in the total spin singlet state, is 
one of the possible channels for multiple exciton generation (MEG).
In chiral single-wall carbon nanotubes (SWCNTs) efficient SF is present within the solar spectrum energy range which is shown by the many-body perturbation theory (MBPT)
calculations based on the density functional theory (DFT) simulations. We 
calculate SF exciton-to-biexction decay rates ${\rm R}_{1\to 2}$ and biexciton-to-exction rates ${\rm R}_{2\to1}$ 
in the (6,2), (6,5), (10,5) SWCNTs, and in (6,2) SWCNT functionalized with Cl atoms. Within the solar energy range, 
we predict ${\rm R}_{1\to2}\sim 10^{14}-10^{15}~s^{-1}$,
while biexciton-to-exction recombination is weak 
with ${\rm R}_{2\to 1}/{\rm R}_{1\to 2}\leq 10^{-2}.$ SF MEG strength in pristine SWCNTs varies strongly with the excitation energy, which is due to highly 
non-uniform density of states at low energy. However, our results for (6,2) SWCNT with chlorine atoms adsorbed to the surface suggest that MEG in the chiral SWCNTs can be
enhanced by altering the low-energy electronic states via surface functionalization.

\end{abstract}

\date{\today}

\maketitle

\section{Introduction}
Increasing the efficiency of photon-to-electron energy conversion in nanomaterials 
has been under active investigation in recent years. 
For instance, one hopes that 
efficiency of the nanomaterial-based solar cells can be increased due to 
carrier multiplication, or multiple exciton generation (MEG) process, where 
absorption of a single energetic photon results in the generation of several excitons \cite{10.1063/1.1736034,ISI:000229120900009,AJ2002115}. 
In the course of MEG the excess photon energy is channeled into creating additional charge carriers 
instead of generating vibrations of the nuclei \cite{AJ2002115}. Indeed, phonon-mediated electron relaxation is a major time evolution channel competing with the MEG. 
The conclusion about MEG efficiency in a nanoparticle can only be made by simultaneously including 
MEG, {phonon-mediated carrier relaxation}, and, possibly, other processes, such as charge and energy transfer \cite{PhysRevB.88.155304,doi:10.1021/jz4004334}.

In the bulk semiconductor materials MEG in the solar photon energy 
range is inefficient \cite{5144200,5014421,10.1063/1.370658}.
In contrast, in nanomaterials MEG is expected to be enhanced by spatial confinement, which increases electrostatic interactions between electrons
\cite{doi:10.1146/annurev.physchem.52.1.193,AJ2002115,doi:10.1021/nl0502672,doi:10.1021/nl100177c,doi:10.1021/ar300189j}. 
%
A potent measure of MEG efficiency is the average number of excitons generated from an absorbed photon -- the  internal quantum efficiency (QE) -- which can be measured in experiments \cite{Semonin16122011}.

MEG has been observed in single-wall carbon nanotubes (SWCNTs) using transient absorption spectroscopy \cite{doi:10.1021/nl100343j} 
and the photocurrent spectroscopy \cite{Gabor11092009};
$QE=1.3$ at the photon energy $\hbar \omega = 3E_g,$ where $E_g$ is the electronic gap, was found in the (6,5) SWCNT.
Theoretically, MEG in SWCNTs has been studied using tight-binding approximation with QE up to 
$1.5$ predicted 
{in (17,0) zigzag SWNT} \cite{PhysRevB.74.121410,PhysRevLett.108.227401}.
{It has been demonstrated that in semiconductor nanostructures MEG is dominated by the impact ionization process \cite{PhysRevLett.106.207401,PhysRevB.86.165319}. 
Therefore, MEG QE requires calculations of the exciton-to-biexciton decay rate (${\rm R}_{1\to2}$) and of the biexciton-to-exciton recombination rate (${\rm R}_{2\to1}$), 
the direct Auger process, and, of course, inclusion of carrier phonon relaxation. In SWCNTs accurate description of these processes requires inclusion of the electron-hole 
bound state effects -- excitons \cite{doi:10.1021/acs.chemrev.5b00012}. 
%
} 
%
%
%

Recently, Density Functional Theory (DFT) combined with the many-body 
perturbation theory (MBPT) techniques has been used to calculate ${\rm R}_{1\to2}$ and ${\rm R}_{2\to1}$ rates, and the photon-to-bi-exciton, 
${\rm R}_2$, and photon-to-exciton, ${\rm R}_1$, rates in two chiral (6,2) and (10,5) SWCNT with different diameters
including exciton effects \cite{doi:10.1063/1.4963735}. QE was then estimated as $QE=({\rm R}_1+2 {\rm R}_2)/({\rm R}_1+{\rm R}_2).$ The results suggested that efficient MEG in chiral SWCNTs
might be present within the solar spectrum range with ${\rm R}_{1\to2}\sim 10^{14}~s^{-1}$,
while ${\rm R}_{2\to1}/{\rm R}_{1\to2}\leq 10^{-2};$ it was found that $QE\simeq 1.2-1.6.$ 
However, MEG strength in these SWCNTs was found to vary strongly with the excitation energy due to highly 
non-uniform density of states. It was suggested that MEG efficiency in these systems could be
enhanced by altering the 
low-energy electronic spectrum via surface functionalization, or simply by mixing SWCNTs of different 
chiralities. 

Another aspect of MEG dynamics has to do with the spin structure of the final bi-exciton state. So far, mostly the simplest possibility of a high-energy spin 
singlet exciton decaying into two spin-zero excitons has been considered in the literature. However, in recent years another possibility for the bi-exciton state 
where a singlet exciton decays into a pair of spin-one exciton states which are in the total spin singlet state -- the singlet fission (SF) -- has received considerable attention. 
(See \cite{doi:10.1021/cr1002613,doi:10.1021/ar300288e} for reviews.) 
This is because triplet excitons tend to have lower energies compared to the singlets and have much longer radiative recombination lifetimes, which may be beneficial for energy conversion 
applications \cite{doi:10.1021/nl070355h}. Also, it has been observed that in some organic molecular crystals, such as various acene and rubrene configurations,
there is resonant energy level alignment between singlet and the double triplet exciton states which enhances SF \cite{C6CE00873A}. 

Properties and dynamics of triplet excitons in SWCNTs have been studied, both experimentally and theoretically \cite{doi:10.1021/nl070355h,PhysRevB.74.121410,nat_phot_2014_s1cnt}. But, to the best of our knowledge, investigation of SF in SWCNTs using DFT-based MBPT has not been attempted.
In this work we develop and apply a DFT-based MBPT technique to explore the possibility of SF in chiral SWCNTs. We calculate ${\rm R}_{1\to 2}$ and ${\rm R}_{2\to 1}$ rates for SF for the (6,2), (6,5), (10,5) SWCNTs, and, also, in (6,2) SWCNT functionalized with Cl atoms. 
{This work aims to provide further insights into the elementary processes contributing to MEG in SWCNTs and its dependence on the chirality, excitation energy, and its sensitivity to the surface functionalization.}

The paper is organized as follows. Section \ref{sec:method} contains description of the methods
and approximations employed in this work. Section III
contains description of the atomistic models studied in this work and of DFT simulation details. 
Section \ref{sec:results} contains discussion of the results obtained. Conclusions and Outlook 
are presented in Section \ref{sec:conclusions}. 
\section{Theoretical Methods and Approximations}
\label{sec:method}
\subsection{Electron Hamiltonian in the KS basis}
\label{sec:H}
The electron field operator $\psi_{\alpha}({\bf x})$ is related to the annihilation operator of the 
$i^{th}$ KS state, ${\rm a}_{i\alpha},$ as
\ber
\psi_{\alpha}({\bf x})=\sum_i\phi_{i\alpha}({\bf x}){\rm a}_{i\alpha},
\label{psi_to_a}
\eer
where $\phi_{i\alpha}({\bf x})$ is the $i^{th}$ KS orbital, and $\alpha$ is the electron spin index \cite{FW,Mahan}. Here we only 
consider spin non-polarzed states with $\phi_{i\uparrow}=\phi_{i\downarrow}\equiv \phi_i;$ also
${\{}{\rm a}_{i\alpha},~{\rm a}_{j\beta}^{\+}{\}}=\delta_{ij}\delta_{\alpha\beta},~{\{}{\rm a}_{i\alpha},~{\rm a}_{j\beta}{\}}=0.$

In the Kohn-Sham (KS) state representation the Hamiltonian of electrons in a CNT is (see, {\it e.g.}, \cite{molphysKK,doi:10.1063/1.4963735})
\ber
{\rm H}=
\sum_{i\alpha}\epsilon_{i} {\rm a}_{i\alpha}^{\+}{\rm a}_{i\alpha}
+{\rm H}_{C}-{\rm H}_{V}+{\rm H}_{e-exciton}.
\label{H}
\eer
where $\epsilon_{i\uparrow}=\epsilon_{i\downarrow}\equiv \epsilon_i$ is the $i^{th}$ KS energy eigenvalue. Typically, in a periodic structure $i={\{}n,{\bf k}{\}},$ where $n$ is the band number, ${\bf k}$ is the lattice wavevector. However, for reasons explained in Section III here KS states are labeled by just integers.
The 
second term is the (microscopic) Coulomb interaction operator 
\ber
{\rm H}_C=\frac12\sum_{ijkl~\alpha,\beta}{\rm V}_{ijkl}{\rm a}^{\dagger}_{i\alpha}{\rm a}^{\dagger}_{j\beta}{\rm a}_{k\beta}{\rm a}_{l\alpha},
~{\rm V}_{ijkl}=\int{\rm d}{\bf x}{\rm d}{\bf y}~\phi^{*}_i({\bf x})\phi^{*}_j({\bf y})\frac{e^2}{|{\bf x}-{\bf y}|}\phi_k({\bf y})
\phi_l({\bf x}).
\label{HC}
\eer
The ${\rm H}_{V}$ term is the compensating potential which prevents double-counting of electron interactions
\ber
{\rm H}_{V}=\sum_{ij}{\rm a}_{i\alpha}^{\+}\left(\int{\rm d}{\bf x}{\rm d}{\bf y}~\phi^*_i({\bf x}){V_{KS}({\bf x},{\bf y})}\phi_j({\bf y})\right){\rm a}_{j\alpha},
\label{HV}
\eer
where $V_{KS}({\bf x},{\bf y})$ is the KS potential consisting of the Hartree and exchange-correlation terms (see, {\it e.g.}, \cite{RevModPhys.74.601,RevModPhys.80.3}). Photon and electron-photon coupling terms are not directly relevant to this work and, so, are not shown, for brevity.

Before discussing ${\rm H}_{e-exciton},$ the last term in the Hamiltonian (\ref{H}),  
let us recall that in the Tamm-Dancoff approximation a spin zero exciton state can be represented as 
\cite{PhysRevB.62.4927,PhysRevB.29.5718}
\ber
\ket{{\alpha}}_{0}={\rm B}^{\alpha\dagger}\ket{g.s.}=\sum_{e h}\sum_{\sigma=\uparrow,\downarrow}\frac{1}{\sqrt{2}} {\rm \Psi}^{\alpha}_{eh} a^{\dagger}_{e\sigma} a_{h\sigma} \ket{g.s.},
\label{Psialpha}
\eer 
where ${\rm \Psi}^{\alpha}_{eh}$ 
is the spin-zero exciton wavefunction, ${\rm B}^{\alpha\dagger}$ is the $\alpha^{th}$ singlet exciton state creation operator; the index ranges are $e>HO,~h\leq HO,$ where HO 
is the highest occupied
KS level, $LU=HO+1$ is the lowest unoccupied KS level.
For a spin one exciton we have
\ber
\ket{{\alpha}}_{1M}={\rm B}_{M}^{\alpha\dagger}\ket{g.s.}=\sum_{e h}\sum_{\mu,\nu} {\rm \Phi}^{\alpha}_{eh} a^{\dagger}_{e\mu} a_{h\nu} {\rm F}^{\mu\nu}_M\ket{g.s.},
~\mu,\nu=\uparrow,\downarrow,
\label{Psialpha1}
\eer 
where ${\rm F}^{\mu\nu}_1=\delta_{\mu\uparrow}\delta_{\nu\downarrow},~{\rm F}^{\mu\nu}_{0}=-{(\sigma_3)_{\mu\nu}}/{\sqrt{2}},~{\rm F}^{\mu\nu}_{-1}=-\delta_{\mu\downarrow}\delta_{\nu\uparrow};~\sigma_i,~i=1,2,3,$ is a Pauli matrix; ${\rm \Phi}^{\alpha}_{eh}$ 
is the spin-one exciton wavefunction, ${\rm B}_{M}^{\alpha\dagger}$ is the triplet exciton creation operator for the state $\alpha$ with spin label $M,~M=-1,0,1.$ Then
\ber
{\rm H}_{e-exciton}&=&
\sum_{e h \alpha}\sum_{\sigma} \frac{1}{\sqrt{2}}\left(\left[\epsilon_{eh}-E^{\alpha}\right]{\rm \Psi}^{\alpha}_{eh}a_{h\sigma}a^{\dagger}_{e\sigma}({\rm B}^{\alpha}+{\rm B}^{\alpha\dagger})+h.c.\right)+\nonumber \\
&&\sum_{e h \alpha}\sum_{\mu \nu}\sum_{M=-1,0,1} \left(\left[\epsilon_{eh}-{\cal E}^{\alpha}\right]{\rm \Phi}^{\alpha}_{eh}a_{h\nu}a^{\dagger}_{e\mu}{\rm F}^{\mu\nu}_M({\rm B}_{M}^{\alpha}+{\rm B}_{M}^{\alpha\dagger})+h.c.\right)+\nonumber \\
&+&\sum_{\alpha}\left(E^{\alpha}{\rm B}^{\alpha\dagger}{\rm B}^{\alpha}+
{\cal E}^{\alpha}\left[\sum_{M=-1,0,1}{\rm B}_{M}^{\alpha\dagger}{\rm B}_{M}^{\alpha}\right]\right),~\epsilon_{eh}=\epsilon_{e}-\epsilon_{h},
\label{Heexc}
\eer
where ${\rm B}^{\alpha\dagger},~E^{\alpha}$ and ${\rm B}_{M}^{\alpha\dagger},~{\cal E}^{\alpha}$ are the singlet and triplet {exciton~creation~operators and energies, respectively.}
The ${\rm H}_{e-exciton}$ term can be seen as the result of, {\it e.g.}, re-summation of perturbative 
corrections to the electron-hole correlation function (see, {\it e.g.}, \cite{Berestetskii:1979aa,Beane:2000fx}); it describes coupling of excitons, both singlets and triplets, to electrons and holes, 
which allows systematic inclusion of excitons in the perturbative calculations 
\cite{PhysRevLett.92.077402,PhysRevLett.92.257402,PhysRevLett.95.247402,Beane:2000fx}.
{To avoid double-counting one chooses the appropriate degrees of freedom, {\it i.e.}, ${\rm a},~{\rm a}^{\dagger}$ or ${\rm B},~{\rm B}^{\dagger},$ which depends on the quantity of interest.}

\begin{figure}
\includegraphics[width=0.96\hsize]{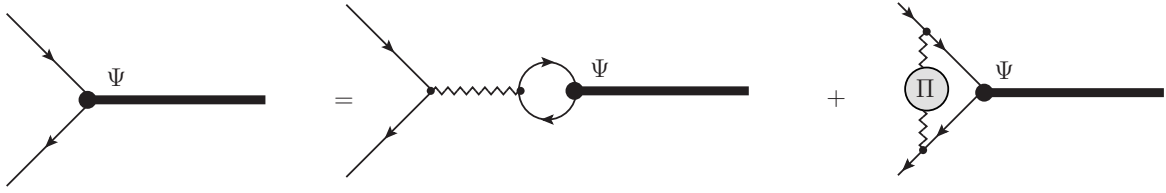}
\caption{Feynman diagrams representing BSE. Thin solid lines represent KS state propagators, thick solid lines are excitons, zigzag lines -- Coulomb potential; ${\Pi}$ is the polarization insertion, 
Eq.~(\ref{Piwkp}).}
\label{fig:BSE}
\end{figure}

To determine
exciton wave functions and energies one solves the Bethe-Salpeter equation (BSE) \cite{PhysRevB.62.4927,PhysRevB.29.5718}.
In the static screening approximation commonly used for semiconductor nanostructures (see, {\it e.g.}, \cite{PhysRevLett.90.127401,PhysRevB.68.085310,PhysRevB.79.245106}) the BSE is \cite{PhysRevB.68.085310}
{
\ber
&&\left([\epsilon_e-\epsilon_h]-E^{\alpha}\right){\rm \Psi}^{\alpha}_{eh}+
\sum_{e^{'}h^{'}} ({\rm c}{\rm K}_{Coul}+{\rm K}_{dir})(e,h;e^{'},h^{'}){\rm \Psi}^{\alpha}_{e^{'},h^{'}}=0,\nonumber \\
&&{\rm K}_{Coul}
=
\sum_{{\bf q}\neq 0}
\frac{8\pi e^2{{\rho}}_{eh}({\bf q}){{\rho}}^{*}_{e^{'}h^{'}}({\bf q})}{V|{\bf q}|^2}, ~
{\rm K}_{dir}
=
-\frac{1}{V}\sum_{{\bf q}\neq 0}
\frac{4\pi e^2{{\rho}}_{e e^{'}}({\bf q}){{\rho}}^{*}_{h h^{'}}({\bf q})}{|{\bf q}|^2-\Pi(0,-{\bf q},{\bf q})},
\label{BSEspin0}
\eer}
where
\ber
{{\rho}}_{ji}({\bf p})=\sum_{{\bf k}}\phi_j^{*}({\bf k}-{\bf p})\phi_i({\bf k}),
\label{rhoij}
\eer
is the transitional density, and
\ber
\Pi(\omega,{\bf k},{\bf p})&=&\frac{8 \pi e^2}{V\hbar}\sum_{ij}\rho_{ij}({\bf k})\rho_{ji}({\bf p})\left(\frac{\theta_{-j}\theta_{i}}
{\omega-\omega_{ij}+i\gamma}-\frac{\theta_{j}\theta_{-i}}{\omega-\omega_{ij}-i\gamma}\right),\nonumber \\
\sum_{i}\theta_i&=&\sum_{i > HO},~\sum_{i}\theta_{-i}=\sum_{i \leq HO},
\label{Piwkp}
\eer
is the RPA polarization insertion (see, {\it e.g.}, \cite{FW}).
Additional screening approximation used in the ${\rm K}_{dir}$ term will be discussed in Section II.~B. 
For the triplet excitons only the direct term contributes, so ${\rm c}=0$ in Eq. (\ref{BSEspin0}) \cite{PhysRevLett.80.3320}.
\vspace{-0.0090095ex}
BSE in terms of the Feynman diagrams is shown in Fig. \ref{fig:BSE}.


In our DFT simulations we have used hybrid Heyd-Scuseria-Ernzerhof (HSE06) exchange correlation functional \cite{vydrov:074106,heyd:219906}, which has been 
successful in reproducing electronic gaps in various semiconductor nanostructures ({\it e.g.}, 
\cite{RevModPhys.80.3,doi:10.1021/ct500958p}). (See, however, \cite{PhysRevLett.107.216806}.) So, here using the HSE06 functional is to 
substitute for $GW$ corrections to the KS energies, {\it i.e.}, for the first step in the standard three-step 
procedure \cite{PhysRevB.34.5390,PhysRevB.62.4927}. Therefore, 
single-particle energy levels and wave functions are approximated by the KS $\epsilon_i$ and 
$\phi_i({\bf x})$ from the HSE06 DFT output. 
While $GW$ technique would improve accuracy of our calculations, it is unlikely to alter our results and conclusions qualitatively. 

Now one is to apply standard perturbative many-body quantum mechanics techniques ({\it e.g.}, \cite{AGD,FW}) to compute the SF decay rates, {\it i.e.}, exciton-to-bi-exciton, bi-exciton-to-exciton rates with the two triplet excitons in the total spin-zero state, working to the second order in the 
screened Coulomb interaction.

As noted above, phonon-meditated electron energy relaxation is an important process competing with MEG. 
A suitable approach to describe time-evolution of a photo-excited nanosystem is the Boltzmann transport equation which includes phonon emission/absorption 
terms together with the terms describing exciton-to-bi-exciton decay and recombination, along with the charge and energy transfer contributions, {\it etc.} 
This challenging task is work in progress.
In this work electron-phonon interaction effects are only included 
by adding small imaginary parts to the KS energies
$\epsilon_i\rightarrow \epsilon_i - i \gamma_i$,
which results in the non-zero line-widths in the expressions below. {In this work all $\gamma$ will be set to 0.025 eV corresponding to room temperature.}

The KS orbital Fourier transformation conventions used in this work are
\ber
&&\phi_i({\bf k}) = \frac{1}{\sqrt{V}}\int_{V} {\rm d}{\bf x}~\phi_i({\bf x}) {\rm e}^{-i{\bf k}\cdot{\bf x}},
~\phi_i({\bf x}) =\frac{1}{\sqrt{V}}\sum_{{\bf k}}\phi_i({\bf k}) {\rm e}^{i{\bf k}\cdot{\bf x}},\nonumber \\
&&{\bf k}
        =
          2\pi
          \left(
                \frac{n_x}{L_x},
                \frac{n_y}{L_y},
                \frac{n_z}{L_z}
          \right),~n_x, n_y, n_z=0,\pm 1, \pm 2,...
\label{phiKS}
\eer
with $V=L_x L_y L_z$ being the simulation cell volume. 

\subsection{Medium Screening Approximation}

For completeness, let us outline the main idea of the simplified treatment of medium screening used in this work \cite{doi:10.1080/00268976.2015.1076580,doi:10.1063/1.4963735}. The standard random phase approximation (RPA) Coulomb potential is
\ber
{\rm W}(\omega,{\bf k},{\bf p})=\frac{4\pi e^2}{V}\left[k^2\delta_{{\bf k},-{\bf p}}-\Pi(\omega,{\bf k},{\bf p})\right]^{-1}.
\label{Wwkp}
\eer
In the static limit $\Pi(\omega,{\bf k},{\bf p})\simeq \Pi(\omega=0,{\bf k},{\bf p}).$ Evaluating ${\rm W}(0,{\bf k},{\bf p})$
requires matrix inversion
which can severely limit applicability of the MBPT techniques \cite{Deslippe20121269,PhysRevB.79.245106}. (See \cite{doi:10.1021/ct500958p} for recent advances.)
In order to be able to simulate nanosystems of interest one is forced to sacrifice some accuracy. With this in mind, a significant technical simplification is to retain 
only the {\it diagonal} matrix elements in $\Pi(0,{\bf k},{\bf p}),$ {\it i.e.}, to approximate
$\Pi(0,{\bf k},{\bf p})\simeq\Pi(0,-{\bf k},{\bf k})\delta_{{\bf k},-{\bf p}}$ as implemented in Eqs. (\ref{BSEspin0},\ref{VC}). In the position space this corresponds to 
$\Pi(0,{\bf x},{\bf x^{'}})\simeq\Pi(0,{\bf x}-{\bf x^{'}}),$
{\it i.e.},
to approximating the system as a uniform medium.
One rationale for this approximation is that in quasi one-dimensional systems, such as CNTs, one can expect $\Pi({\bf x},{\bf x^{'}})\simeq\Pi(z-z^{'}),$ 
where $z,z^{'}$ are the axial positions. 

Previously, we have checked quality of our computational approach including this screening approximation for chiral SWCNTs \cite{doi:10.1063/1.4963735}.
We have computed low-energy absorption spectra for (6,2) and (10,5) SWCNTs and found that 
our predictions for $E_{11}$ and $E_{22}$ -- the energies of the first two absorption peaks corresponding to
transitions between the van Hove peaks in the CNT density of states -- reproduce results of Weisman and Bachillo \cite{doi:10.1021/nl034428i} 
within {5 - 13 \% error.} Additionally, we have simulated SWCNT (6,5) and found $E_{11}=1.1~eV,~E_{22}=2.05~eV$ {\it vs.} $E_{11}=1.27~eV~eV,~E_{22}=2.19~eV$ from \cite{doi:10.1021/nl034428i}.
{This suggests that our approach is adequate for the semi-quantitative description of these systems.}
Accuracy could be improved by using full interaction ${\rm W}(0,{\bf k},{\bf p}),$ or ${\rm W}(\omega,{\bf k},{\bf p})$, and GW, which would be much more computationally 
expensive.  
However, it would not 
change the overall conclusions of this work.

\subsection{Expressions for the Rates}
\begin{figure}[!t]
\center
\includegraphics*[width=0.995\textwidth]{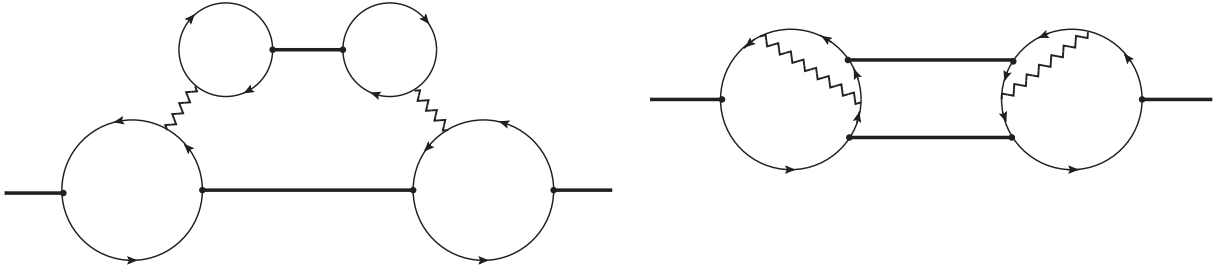}
\caption{Exciton self-energy Feynman diagrams for the exciton$\to$bi-exciton process. Thin solid lines stand for the KS state propagators, thick solid lines depict excitons, zigzag lines -- screened Coulomb potential. 
The diagrams on the left and the right correspond to the exchange and direct channels, respectively. Not shown for brevity are the similar diagrams with all the Fermion arrows reversed. Only the direct channel diagram contributes to SF. For SF final bi-exciton state is understood to be the singlet.
}
\vspace{-0.45ex}
\label{fig:R1to2}
\end{figure}

Within our approximations exciton-to-bi-exciton decay rate from the impact ionization process is given by
\ber
{\rm R}_{1{\rightarrow}2}=-2{\rm Im}\Sigma_{\gamma}(\omega_{\gamma}), 
\label{R1to2}
\eer
where 
$\Sigma_{\gamma}(\omega)$ are the exciton-to-bi-exciton decay contributions to the self-energy function of 
the exciton state $\gamma$ with energy $E^{\gamma}=\hbar\omega_{\gamma}.$ 
The relevant self-energy Feynman diagrams are shown in Fig. \ref{fig:R1to2}.

For completeness, let us quote the expressions for the all-singlet exciton-to-bi-exciton rates \cite{doi:10.1063/1.4963735}
\ber
R_{1{\to}2}(\omega_{\gamma})&=&R^p+R^h+{\tilde R}^p+{\tilde R}^h,\nonumber \\
 R^p(\omega_{\gamma})&=&2\frac{2 \pi}{\hbar^2}\sum_{\alpha\beta}\delta(\omega_{\gamma}-\omega_{\alpha} -\omega_{\beta})
\abs*{\sum_{ijkln}W_{{jlnk}}
\theta_l \theta_{-n} 
(\Psi_{ln }^{\beta }){}^*
\theta_i \theta_{-j} \theta_{-k}
\Psi_{{ij}}^{\gamma} 
\left(\Psi_{{ik}}^{\alpha }\right){}^*
}^2,\nonumber \\
R^h(\omega_{\gamma})&=&2\frac{2 \pi}{\hbar^2}\sum_{\alpha\beta}\delta(\omega_{\gamma}-\omega_{\alpha} -\omega_{\beta})
\abs*{\sum_{ijkln}W_{{jlnk}} 
\theta_{-l} \theta_n 
\Psi_{{nl}}^{\beta }
\theta _{-i} \theta_j \theta _k
   (\Psi_{{ji}}^{\gamma }){}^* \Psi_{{ki}}^{\alpha}
}^2.
\label{R1to2all}
\eer
The expressions for ${\tilde R}^h$ and ${\tilde R}^p$ are the same as the ones for $R^h,~R^p$ with $W_{jlnk}$ replaced by $W_{jlkn}$ and divided by 2.

A spin-singlet state composed of two noninteracting spin-one excitons is ({\it cf.} Eq. 5 of \cite{doi:10.1063/1.4794425})
\ber
\ket{\alpha\beta}_{TT;0}&=&\frac{1}{\sqrt{3}}\left({\rm B}_{1}^{\alpha\dagger}{\rm B}_{-1}^{\beta\dagger}-{\rm B}_{0}^{\alpha\dagger}{\rm B}_{0}^{\beta\dagger}+{\rm B}_{-1}^{\alpha\dagger}
{\rm B}_{1}^{\beta\dagger}\right)\ket{g.s.}=\nonumber \\ &=&\sum_{e,h,e^{'},h^{'}} \sum_{\mu,\nu,\lambda,\sigma}{\rm T}^{\mu\nu\lambda\sigma}{\rm \Phi}^{\alpha}_{eh}{\rm \Phi}^{\beta}_{e^{'}h^{'}} a^{\dagger}_{e\mu} a_{h\nu} 
a^{\dagger}_{e^{'}\lambda} a_{h^{'}\sigma}\ket{g.s.},\nonumber \\
{\rm T}^{\mu\nu\lambda\sigma}&=&-\frac{1}{\sqrt{3}}\left(\delta_{\mu\sigma}\delta_{\nu\lambda}-\frac{1}{2}\delta_{\mu\nu}\delta_{\lambda\sigma}\right).
\label{2t00}
\eer
 
The expressions for the singlet fission rate, {\it i.e.}, the rate for the singlet-to-two-triplets process, are
\ber
R^{SF}_{1{\to}2}(\omega_{\gamma})&=&{\rm R}^p+{\rm R}^h,\nonumber \\
 {\rm R}^p(\omega_{\gamma})&=&\frac{2 \pi}{\hbar^2}\frac{3}{2}\sum_{\alpha\beta}\delta(\omega_{\gamma}-\omega_{1,\alpha} -\omega_{1,\beta})
\abs*{\sum_{ijkln}W_{{jlkn}}
\theta_l \theta_{-n} 
(\Phi_{ln }^{\beta }){}^*
\theta_i \theta_{-j} \theta_{-k}
\Psi_{{ij}}^{\gamma} 
\left(\Phi_{{ik}}^{\alpha }\right){}^*
}^2,\nonumber \\
{\rm R}^h(\omega_{\gamma})&=&\frac{2 \pi}{\hbar^2}\frac{3}{2}\sum_{\alpha\beta}\delta(\omega_{\gamma}-\omega_{1,\alpha} -\omega_{1,\beta})
\abs*{\sum_{ijkln}W_{{jlkn}} 
\theta_{-l} \theta_n 
\Phi_{{nl}}^{\beta }
\theta _{-i} \theta_j \theta _k
   (\Psi_{{ji}}^{\gamma }){}^* \Phi_{{ki}}^{\alpha}
}^2,
\label{R1to2triplet}
\eer
where ${\cal E}^{\gamma}=\hbar\omega_{1,\gamma}.$ 
In the above
\ber
W_{jlnk}&=&\sum_{{\bf q}\neq 0}\frac{4 \pi e^2}{V}\frac{{{\rho}}_{kj}^{*}({\bf q}){{\rho}}_{ln}({\bf q})}
{\left(q^2-\Pi(0,-{\bf q},{\bf q})\right)}
\label{VC}
\eer
is the (approximate) screened Coulomb matrix element, and
\ber
\delta(x)=\frac1\pi\frac{\gamma}{x^2+\gamma^2},
\label{delta_L}
\eer
the Lorentzian representation of the $\delta$-function. Only the direct channel diagram (Fig. \ref{fig:R1to2}, on the right) contributes to SF.

In the above expressions only the terms leading in 
the ratio of the typical exciton binding energy to the HO-LU gap $\epsilon_{binding}/E_g < 1$ are shown, for brevity. 

The rate as a function of energy is given by averaging over the initial exciton states within given energy range with the 
$\gamma=0.025~eV$ resolution, {\it i.e.},
\ber
R(\epsilon)=\frac{1}{N(\epsilon)}\sum_{\alpha}R(E^{\alpha}),
\label{R1to2ave}
\eer
where the sum is over the exciton states within the $(\epsilon,\epsilon+\gamma)$ energy range, $N(\epsilon)$ is the number of such states.

The above expressions have the overall structure of the Fermi Golden Rule.   
The bi-exciton-to-exciton rate expressions are given by similar expressions 
with the initial and final states reversed.

{\section{Computational Details}}
\label{sec:compdetail}
The optimized geometries and KS orbitals and KS energy eigenvalues of the chiral SWCNTs studied here 
have been obtained using the {\it{ab initio}} total energy and
molecular dynamics program VASP (Vienna ab initio simulation program) 
with the hybrid Heyd-Scuseria-Ernzerhof (HSE06) exchange correlation functional \cite{vydrov:074106,heyd:219906}
using the 
projector augmented-wave (PAW) pseudopotentials \cite{PhysRevB.50.17953,PhysRevB.59.1758}.
\begin{figure}[!t]
\center
\vspace{-8.5cm}
\includegraphics*[width=0.995\textwidth]{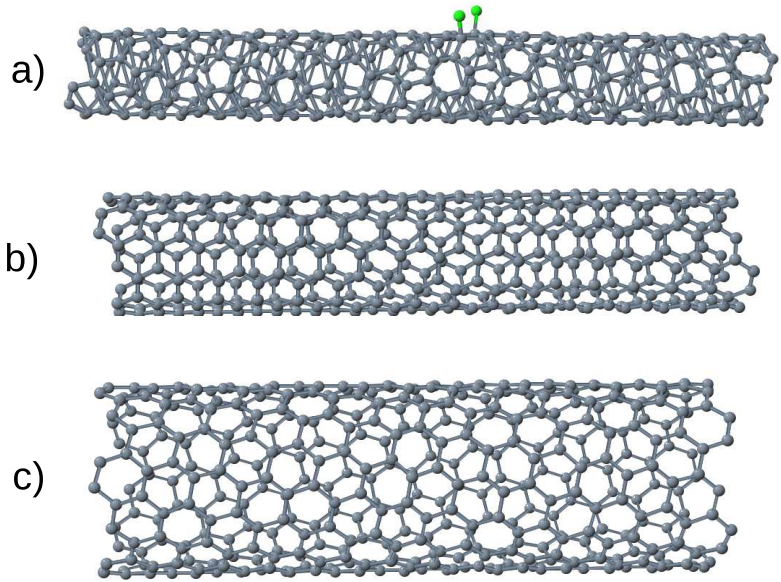}
\vspace{-7cm}
\caption{Atomistic models of chiral SWCNTs. Shown in a) is (6,2) with two chlorine atoms adsorbed to the surface in a para configuration. In order to keep the doping concentration low three unit cells have been included in the simulations. In b) is SWCNT (6,5). Only one unit cell is included due to computational cost restrictions. In c) is (10,5) with three unit cells.
}
\vspace{-0.45ex}
\label{fig:optimized_structures}
\end{figure}
Using conjugated gradient method for ion position relaxation the 
structures were relaxed until residual forces on the ions 
were no greater than $0.05~eV/\AA.$ 
The momentum cutoff defined by
\ber
\frac{\hbar^2{k}^2}{2 m}\leq {\cal E}_{max},
\label{Ecutoff}
\eer
where 
$m$ is the electron mass, was set to ${\cal E}_{max}=400~eV.$
The number of KS orbitals included in the simulations which regulated energy cutoff
were chosen so that 
$\epsilon_{i_{max}}-\epsilon_{HO}\simeq\epsilon_{LU}-\epsilon_{i_{min}}\geq 3~eV,$ 
where $i_{max},~i_{min}$ are the highest and the lowest KS labels included in simulations.

SWCNT atomistic models 
were placed in various finite volume simulation boxes with periodic boundary conditions where in the axial direction the length of the box has been chosen to accommodate 
an integer number of unit cells, while in the other two directions the SWCNTs 
have been kept separated by about $1~nm$ of vacuum surface-to-surface thus 
excluding spurious interactions between their periodic images.

Previously, we have found reasonably small (about 10\%) variation in the single 
particle energies over the Brillouin zone when three unit cells were included in the DFT simulations \cite{doi:10.1063/1.4963735}.
So, simulations have been done including three unit cells of (6,2) and (10,5) SWCNTs 
at the $\Gamma$ point.
So, in our approximation lattice momenta of the KS states, which are suppressed by the reduced Brillouin zone size, have been neglected. For (6,5) SWCNT due to high computational cost only 
one unit cell was included. But as mentioned above, simulation based on this size-reduced model reproduced the absorption spectrum features with the same accuracy as other SWCNTs. (See Table I.)
 
The rationale for including more unit cells instead of standard sampling of the Brillouin zone by including more $K$-points in the DFT simulations is that surfaces of these SWCNTs are to be functionalized.
Inclusion of several unit cells allows us to keep the concentration of surface 
dopants reasonably low. So, here we have simulated (6,2) SWCNT doped with chlorine, where two $Cl$ atoms are attached to the same carbon ring in the para configuration, which has been found to be the 
preferred arrangement 
\footnote{Private communication with S.~Kilina.} 

The atomistic models of the optimized nanotubes are shown in Fig. (\ref{fig:optimized_structures}).
%
%
In this work all the DFT simulations have been done in a vacuum which should be adequate to describe properties of these SWCNTs 
dispersed in a non-polar solvent. 

\section{Results and Discussion}
\label{sec:results}
\begin{figure}[!t]
\center
\begin{tabular}{cc}
\vspace{-5.25ex}
\raisebox{8.175\totalheight}{\hspace{-4.25ex} (a)}
\raisebox{0.185\totalheight}{\includegraphics*[width=0.465\textwidth] 
{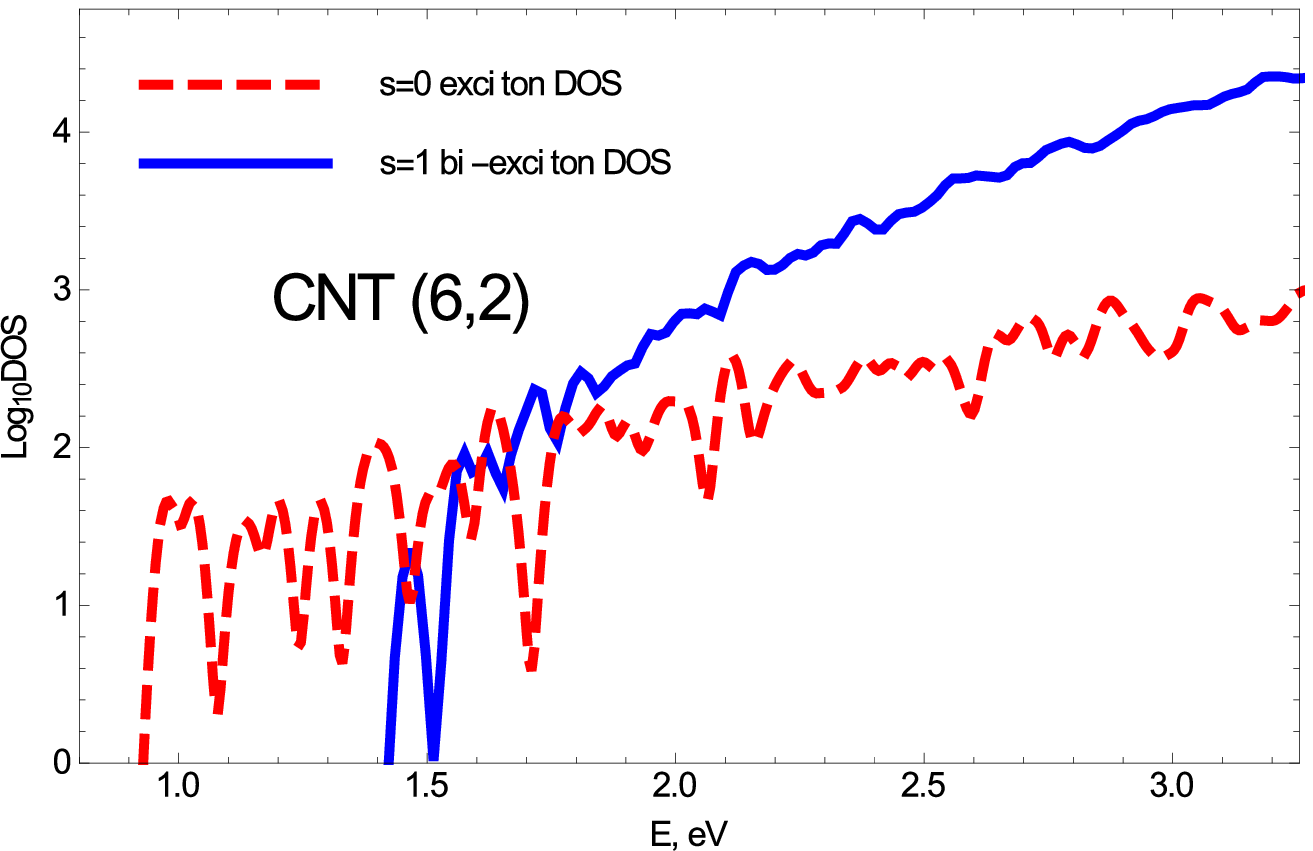}} &
\raisebox{8.175\totalheight}{(b)}\raisebox{0.175\totalheight}{\includegraphics*[width=0.495\textwidth] 
{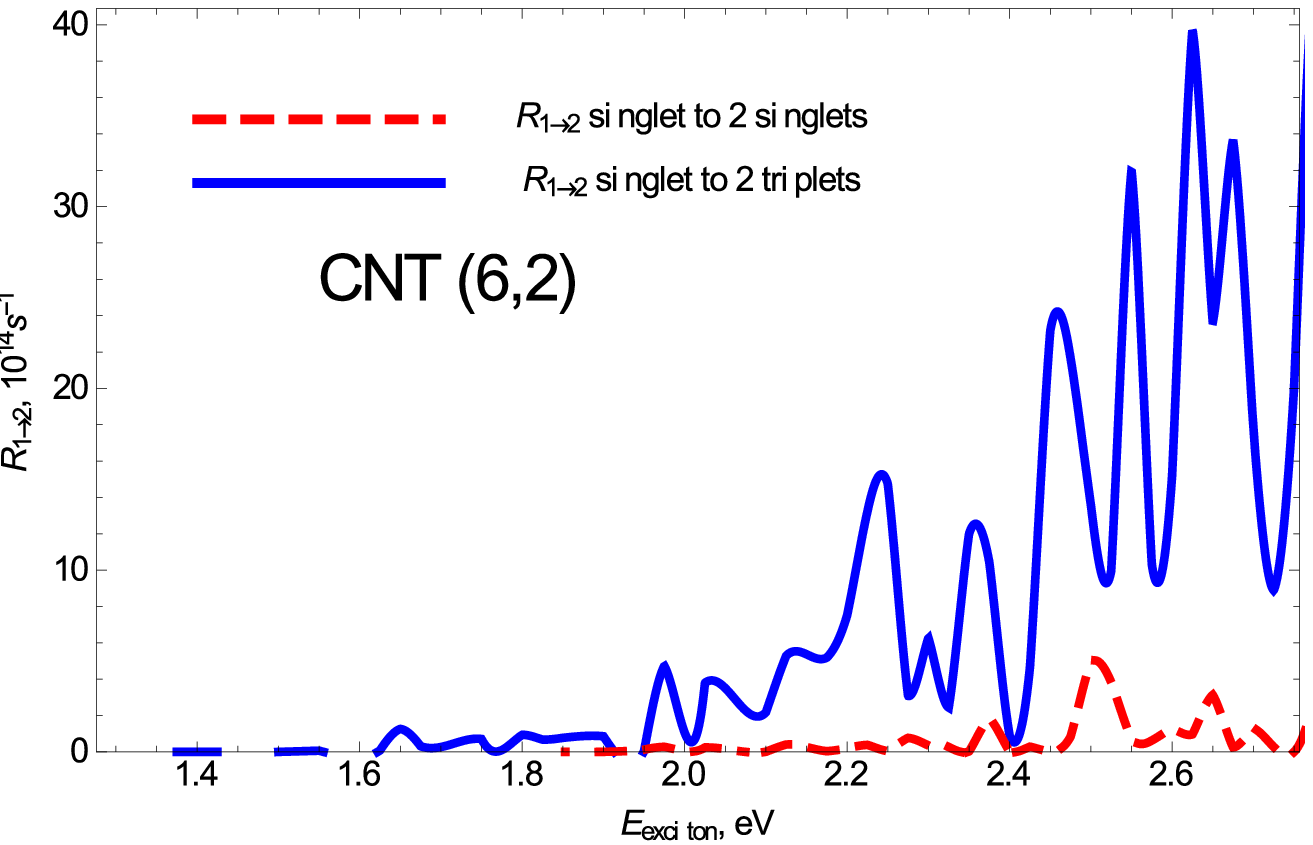}}\\
\vspace{-2.25ex}
\raisebox{9.175\totalheight}{\hspace{-3.25ex} (c)}
\raisebox{0.195\totalheight}{\includegraphics*[width=0.465\textwidth] 
{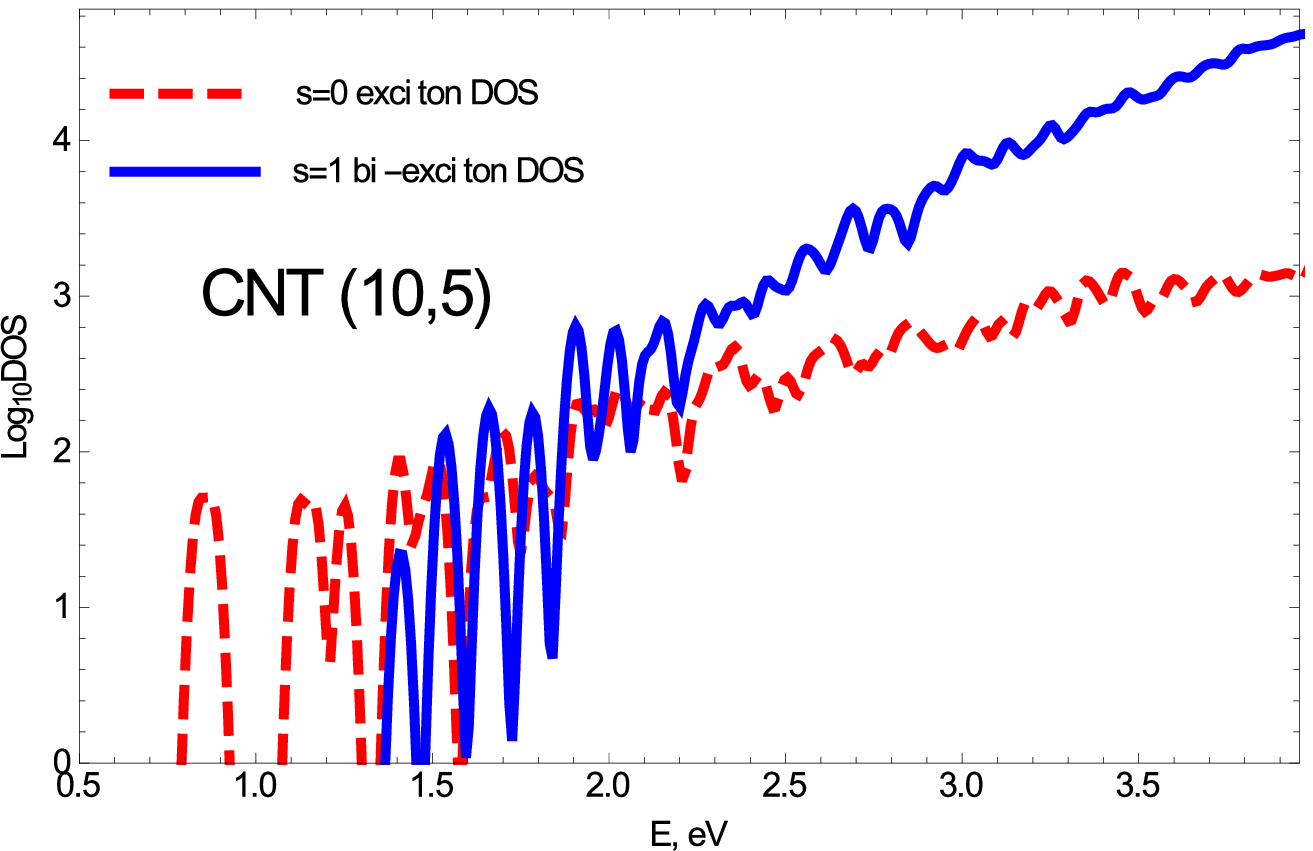}}&
\raisebox{9.175\totalheight}{\hspace{-2.25ex} (d)}
\raisebox{0.195\totalheight}{\includegraphics*[width=0.485\textwidth] 
{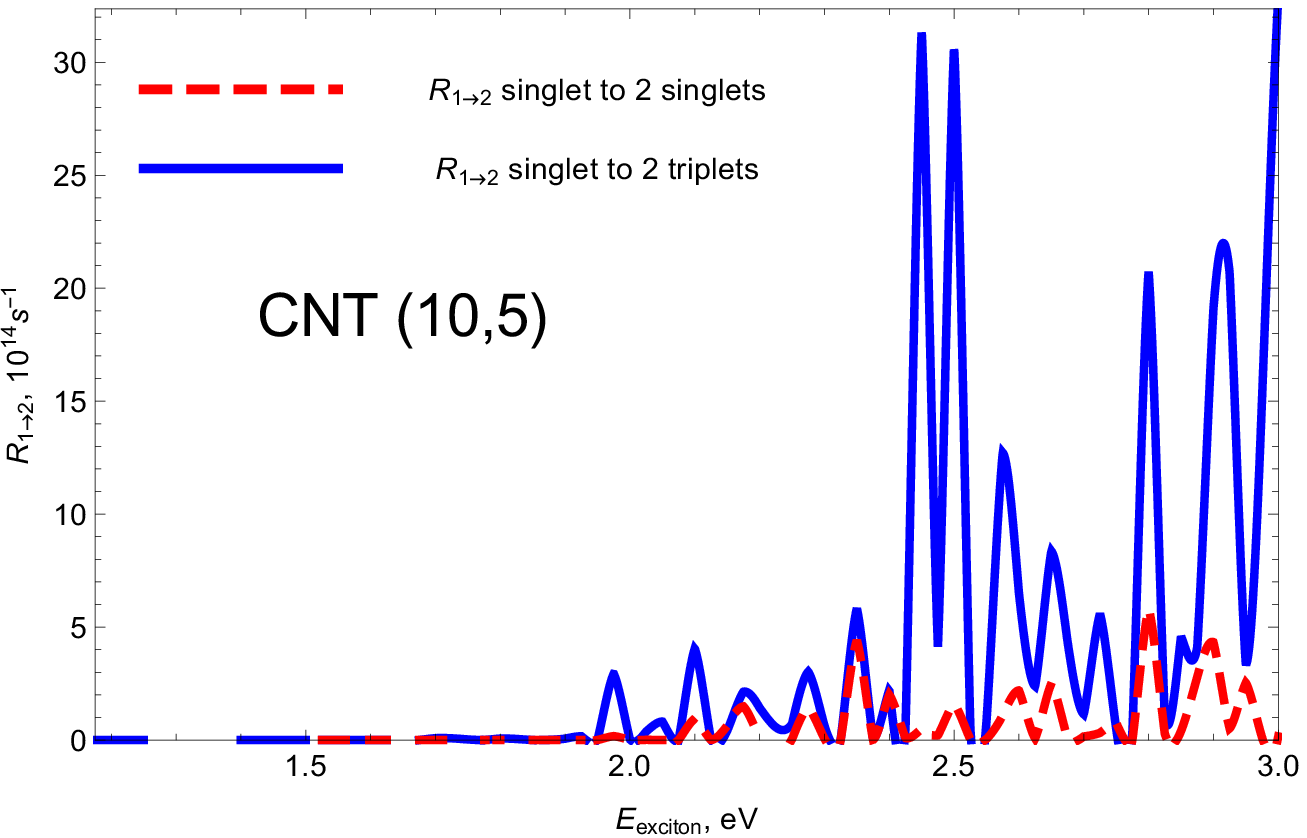}}\\
\vspace{-2.25ex}
\raisebox{9.175\totalheight}{\hspace{-3.25ex} (e)}
\raisebox{0.195\totalheight}{\includegraphics*[width=0.465\textwidth] 
{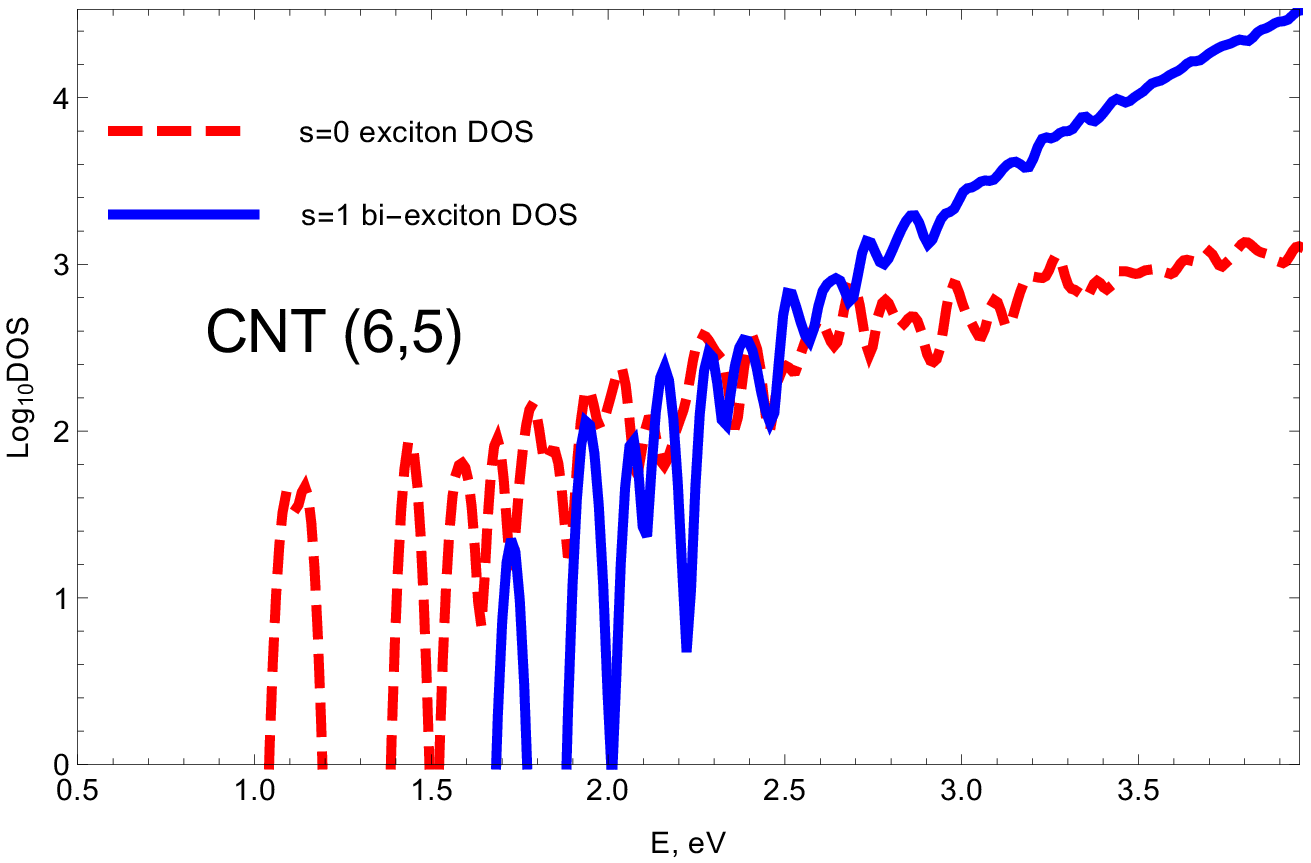}}&
\raisebox{9.175\totalheight}{\hspace{-2.25ex} (f)}
\raisebox{0.195\totalheight}{\includegraphics*[width=0.485\textwidth] 
{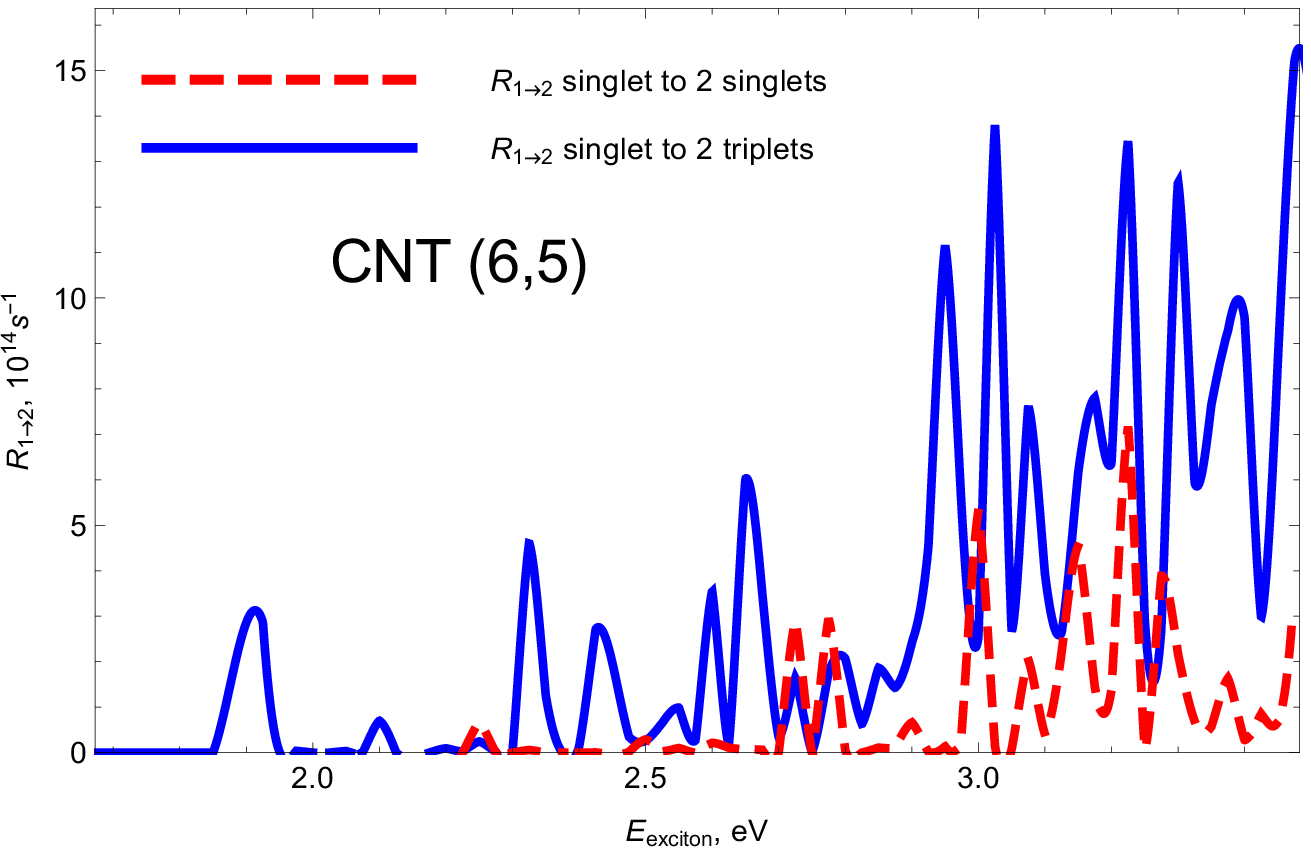}}\\
\end{tabular}
\caption{Singlet exciton and triplet biexciton densities of states (DOS) 
and the MEG $R_{1\to 2}$ rates, all-singlet and SF, for the (6,2) ((a) and (b)), (10,5) ((c) and (d)) and (6,5) ((e) and (f)) CNTs. The rates for (6,2) and (10,5) are from \cite{doi:10.1063/1.4963735} and shown here for comparison. (Color on-line only.)} 

\vspace{-0.25ex}
\label{fig:excDOS_R12_62_105_65}
\end{figure}
\begin{figure}[!t]
\center
\begin{tabular}{cc}
\vspace{-5.25ex}
\raisebox{8.175\totalheight}{\hspace{-4.25ex} (a)}
\raisebox{0.185\totalheight}{\includegraphics*[width=0.465\textwidth] 
{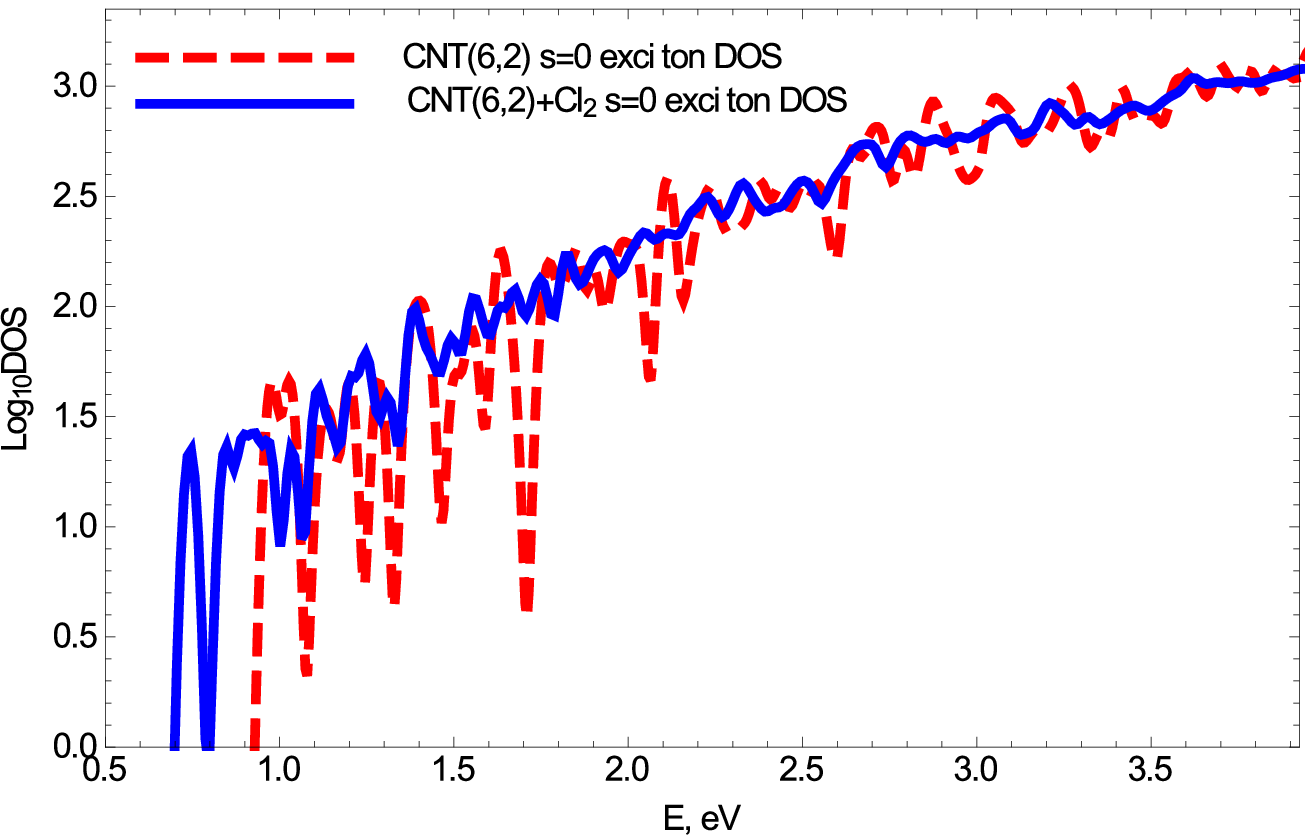}} &
\raisebox{8.175\totalheight}{(b)}\raisebox{0.17\totalheight}{\includegraphics*[width=0.485\textwidth] 
{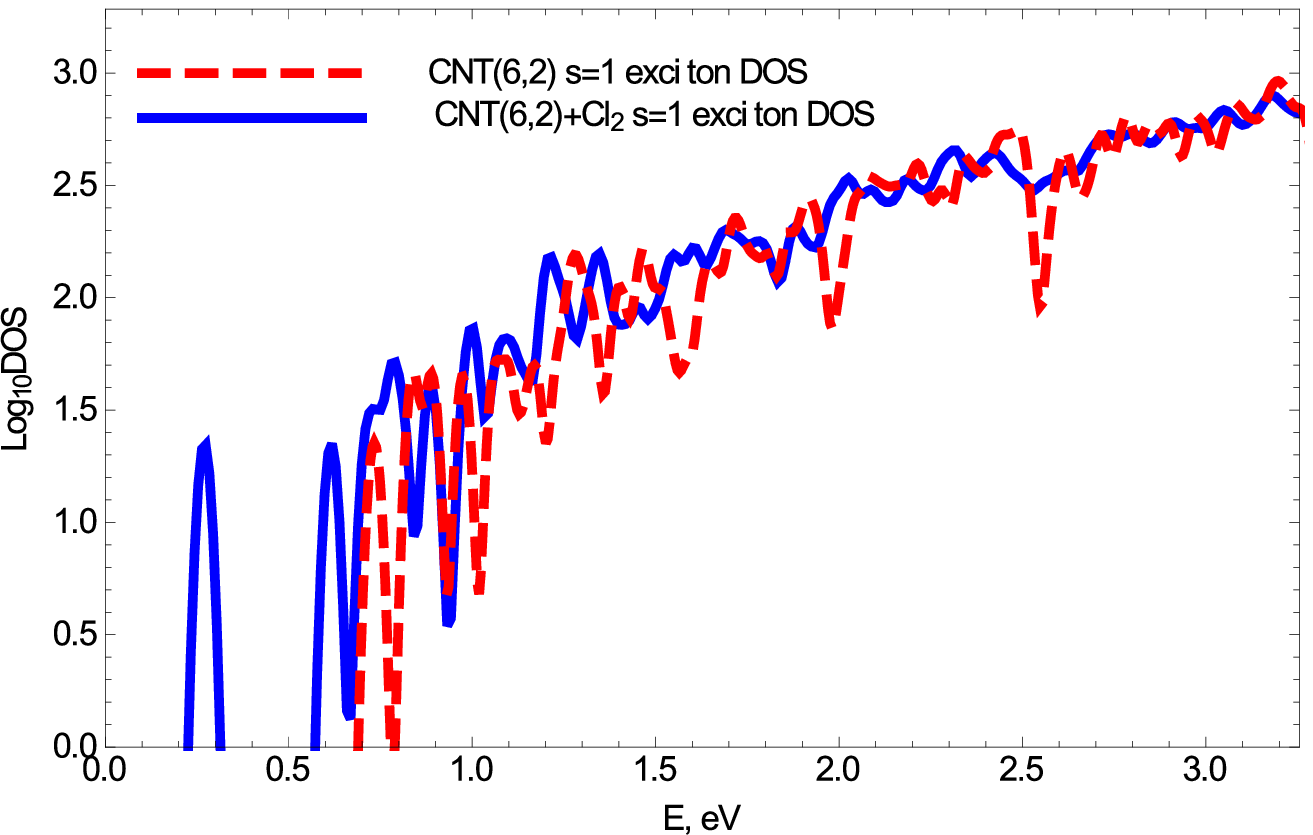}}\\
\vspace{-2.25ex}
\raisebox{9.175\totalheight}{\hspace{-3.25ex} (c)}
\raisebox{0.195\totalheight}{\includegraphics*[width=0.465\textwidth] 
{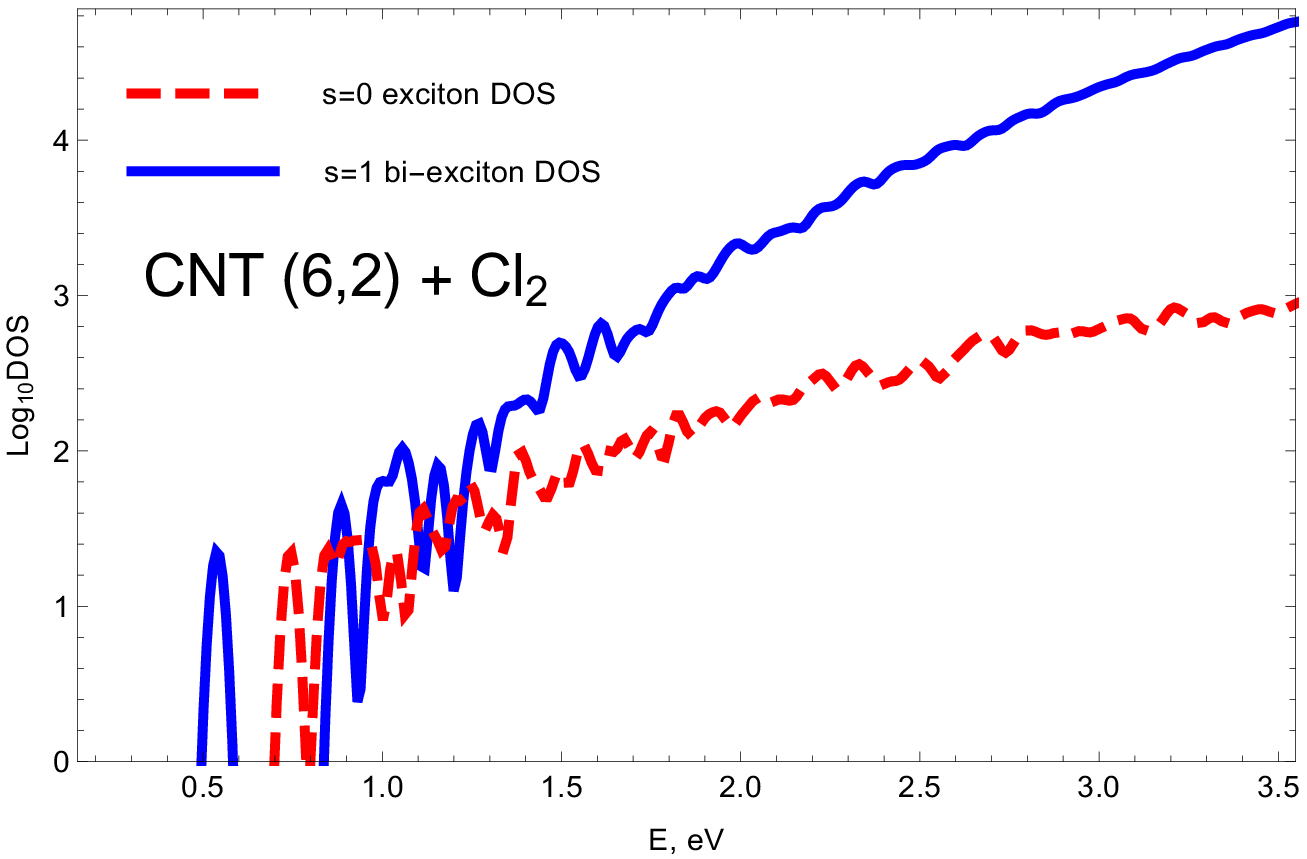}}&
\raisebox{9.175\totalheight}{\hspace{-0.25ex} (d)}
\raisebox{0.2095\totalheight}{\includegraphics*[width=0.485\textwidth] 
{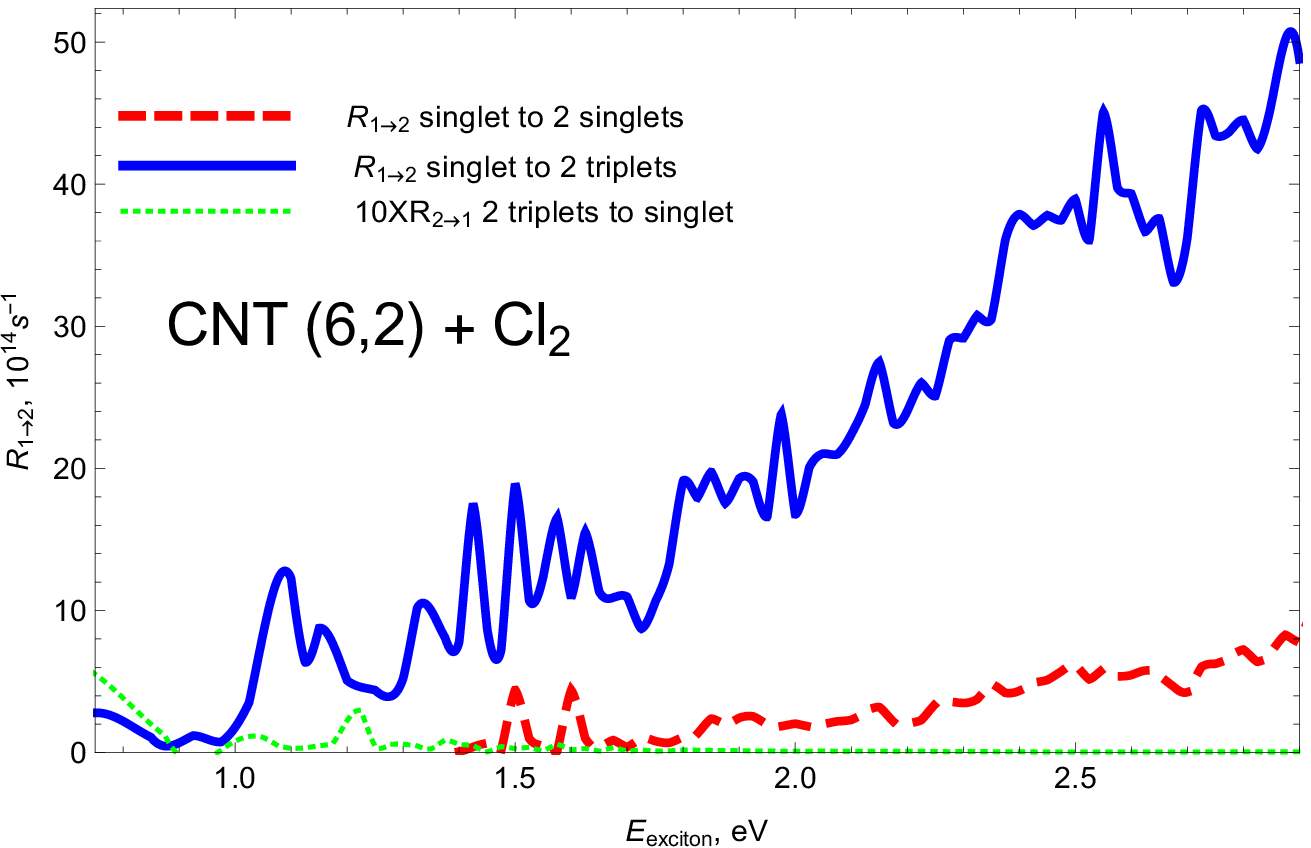}}\\
\end{tabular}
\caption{Exciton DOS and MEG rates for the pristine and doped (6,2) SWCNT. Shown in (a) are the singlet exciton DOSs for the pristine and $Cl$ doped (6,2) SWCNT; in (b) -- the triplet exciton DOSs for the pristine and doped (6,2) SWCNT. Shown in (c) are the singlet exciton and triplet biexciton DOSs for the doped (6,2) SWCNT. In (d) are the MEG rates for the $Cl$ doped (6,2) SWCNT: dashed (red) line depicts the all-singlet exciton-to-biexciton rate $R_{1\to 2}$, solid (blue) line -- the SF exciton-to-biexciton rate. The (green) dotted line corresponds to the biexciton-to-exciton rate $R_{2\to1}$ of the $Cl$ doped (6,2) SWCNT. {\bf $R_{2\to1}$ has been multiplied by 10 for better presentation .} This recombination rate is the greatest of all the cases considered here. (Color on-line only.)
}
\vspace{-0.25ex}
\label{fig:R_62Cl}
\end{figure}

The main results are shown in Table I and in Figs. (\ref{fig:excDOS_R12_62_105_65}),~(\ref{fig:R_62Cl}). 
We have found (see Table I) that in all cases the lowest triplet exciton energy is red-shifted 
compared to the singlet, which is as expected since the repulsive exchange contribution to the 
BSE kernel is absent for the triplets \cite{doi:10.1021/nl070355h}. As a result, the energy threshold for SF is somewhat lower 
compared to the all-singlet MEG. The SF and all-singlet MEG rates for pristine (6,2), (10,5) and (6,5) SWCNTs 
are shown in Fig. (\ref{fig:excDOS_R12_62_105_65}). Shown here for comparison are the all-singlet rates for (6,2) and (10,5) are from \cite{doi:10.1063/1.4963735}.
{\begin{table}
\raisebox{0.00001\totalheight}{\begin{tabular}{|c|c|c|c|c|}\hline
\textbf{} & $(6,2)$ & $(6,2)+Cl_2$ & $(6,5)$ & $(10,5)$ \\ \hline
\textbf{$E_g,~eV$} & 1.33 & 0.96 & 1.22 & 0.91 \\ \hline
\textbf{$E_g^{BSE}~s=0,~eV$} & 0.98 & 0.74 & 1.09 & 0.835 \\ \hline
\textbf{$E_g^{BSE}~s=1,~eV$} & 0.73 & 0.27 & 0.86 & 0.71 \\ \hline
\end{tabular}}
\label{Pik-kresults}
\caption{
$E_g\equiv\epsilon_{LU}-\epsilon_{HO}$, 
is the HO-LU gap, $E^{BSE}_g$ is the minimal exciton energy from BSE for the singlets ($s=0$) and triplets ($s=1$).
} 
\end{table}}
Our calculations predict that efficient MEG both in the SF and all-singlet channels is present in chiral SWCNTs within the solar spectrum range but 
its strength varies strongly with the excitation energy. This is clearly due to the highly non-uniform low-energy 
electronic spectrum in SWCNTs (see Fig. \ref{fig:excDOS_R12_62_105_65}, (a), (c), (e)). The $R_{1\to 2}$ MEG rates reach $10^{14}-10^{15}~1/s$ (see Fig. \ref{fig:excDOS_R12_62_105_65}, (b), (d), (f)). The recombination rates $R_{2\to1}$ are 
suppressed for all energies with $R_{2\to 1}/R_{1\to 2} \leq 10^{-2}$ \cite{doi:10.1063/1.4963735}; they are not shown. In (6,2) the all-singlet MEG starts at the energy threshold 
$2 \times E_g=1.95~eV,$ the SF -- at $2.3 \times E^t_g=1.7~eV,$ where 
$E^t_g$ is the minimal triplet exciton energy, but in (10,5) the all-singlet MEG becomes appreciable at about $2.4 \times E_g=2.0~eV;$ the threshold for SF is $2.75 \times E^{t}_g=1.95~eV.$ In (6,5) the all-singlet MEG starts at $2.1 \times E_g=2.25~eV,$ SF -- at $2.2 \times E^t_g=1.9~eV,$

Shown in Fig. \ref{fig:R_62Cl} are results for 
the (6,2) SWCNT with chlorine atoms attached to the surface as described in Section \ref{sec:compdetail}. 
Complete discussion of the influence of this surface defect on the system's optoelectronic properties will be presented elsewhere. 
As far as the MEG-related properties are concerned, we predict that doping significantly red-shifts exciton energy spectra, both singlet (Fig. \ref{fig:R_62Cl}, (a)) and triplet (Fig. \ref{fig:R_62Cl}, (b)). 
DOS for the initial and final MEG states are shown in Fig. \ref{fig:R_62Cl}, (c). In this case, SF MEG is energetically allowed even for the lowest singlet exciton. 
Shown in Fig. \ref{fig:R_62Cl}, (d) are the MEG rates for the $Cl$-decorated (6,2) SWCNT. The all-singlet MEG threshold is at about $2 E_g=1.5~eV;$ the threshold for SF is $0.75~eV,$ which is the lowest singlet exciton energy. Importantly, both the all-singlet and SF MEG rates $R_{1 \to 2}$ are much less oscillatory as a function of the exciton energy than the pristine case rates ({\it cf.} Fig. \ref{fig:excDOS_R12_62_105_65}, (b) and \ref{fig:R_62Cl}, (d)). The recombination rate $R_{2\to 1}$ -- which is the greatest of all the cases considered -- is shown in Fig. \ref{fig:R_62Cl}, (d). Note that it is multiplied by 10 for better presentation.

In all cases we find that SF rates are greater in magnitude than the all-singlet rates. This is likely due to the aforementioned overall 
red-shift of the triplet biexciton spectrum compared to the singlet exciton energies. While the Coulomb interaction matrix elements between 
the electron/hole and trion states are similar in magnitude in both cases, for the same energy there are simply more available bi-exciton final states 
for the SF than for the all-singlet channel.
\section{Conclusions and Outlook}
\label{sec:conclusions}

Working to the second order in the screened Coulomb interaction 
and including electron-hole bound state effects we have developed a DFT-based 
MBPT technique for SF which allows one to compute the exciton-to-bi-exciton and 
the inverse bi-exciton-to-exciton rates when the initial state is a high-energy singlet
while the final state is a pair of non-interacting triplet excitons in spin-correlated state with the
total spin zero. Then, this method was used to calculate MEG in the chiral SWCNTs, using (6,2), (6,5) and (10,5) as examples. Also, we 
have simulated (6,2) SWCNT with chlorine atoms adsorbed to the surface.

Our calculations suggest that chiral SWCNTs have efficient MEG within 
the solar spectrum range both for the all-singlet channel and SF with 
$R_{1\to 2} \sim 10^{14}-10^{15}~s^{-1}$ and with the 
recombination rates suppressed as $R_{2\to 1}/R_{1\to 2} \sim 10^{-2}.$ In the pristine SWCNTs the
MEG rates vary strongly with the excitation energy. In contrast, our results for the $Cl$-decorated (6,2) SWCNT
suggest that surface functionalization significantly alters low-energy spectrum in a SWCNT. As is typical for doping,
the defect creates additional shallow electronic states, which improves MEG efficiency. In the doped case, $R_{1\to 2}$ is not 
only greater in magnitude, but also is a much smoother function of the excitation energy.  
An alternative way to increase efficiency of carrier multiplication is to use SWCNT mixtures 
of different chiralities. 

As noted above, an investigation of MEG efficiency in a nanosystem should be comprehensive, {\it i.e.},
carrier multiplication and biexciton recombination should be allowed to ``compete" with other processes, such as
phonon-mediated carrier relaxation, energy and charge transfer, {\it etc.} \cite{doi:10.1021/jz4004334}. The Kadanoff-Baym-Keldysh, or NEGF,
technique is a suitable formalism to achieve this goal \cite{Landau10,PhysRevB.83.165306,PhysRevLett.112.257402}. Bi-exciton creation and recombination, both in the all-singlet and SF channels, 
phonon emission, recombination, energy and charge transfer and other effects are to be included in the transport equation describing time evolution of a weakly non-equilibrium photoexcited state.

As described above (see Section II), our calculations had to utilize several simplifying approximations. However, we have verified that 
our results for the absorption spectra are in reasonable agreement with experimental data with the error less then 13\% for E$_{11}$ and E$_{22}$ excitonic bands 
for the (6,2), (6,5) and (10,5) nanotubes. This suggests overall applicability of our technique for these systems
at least at the semi-quantitative level.  
Accuracy of our methods can be further improved in several ways. One natural improvement is to calculate $GW$ 
single particle energy corrections, which then can be easily incorporated in the 
rate expressions. {It is likely to blue-shift the rate curves 
by a fraction of eV without significant changes to the shape.} Another step is to use full RPA 
interaction ${\rm W}(0,{\bf k},{\bf p})$ rather than ${\rm W}(0,-{\bf k},{\bf k}).$ 
Also, in the impact ionization process the typical energy exchange exceeds the gap and, so, role 
of dynamical screening needs to be investigated. 
Going beyond second order in the screened Coulomb interaction would require keeping the wave function renormalization factor 
(see, {\it e.g.}, \cite{FW}) in the exciton decay rate expressions in Eqs. (\ref{R1to2}), (\ref{R1to2triplet}).
However, none of these corrections are likely to change the main results of this work,
while drastically increasing computational cost.

\section{Acknowledgments}
Authors acknowledge financial support from the NSF grant CHE-1413614. 
The authors acknowledge the use of computational resources at
the Center for Computationally Assisted Science and
Technology (CCAST) at North Dakota State University and
the National Energy Research Scientific Computing Center
(NERSC) allocation award 86678, supported by the Office of
Science of the DOE under contract No. DE-AC02-05CH11231.







%
%
\bibliography{dftNSF2013}

\begin{thebibliography}{57}
\expandafter\ifx\csname natexlab\endcsname\relax\def\natexlab#1{#1}\fi
\expandafter\ifx\csname bibnamefont\endcsname\relax
  \def\bibnamefont#1{#1}\fi
\expandafter\ifx\csname bibfnamefont\endcsname\relax
  \def\bibfnamefont#1{#1}\fi
\expandafter\ifx\csname citenamefont\endcsname\relax
  \def\citenamefont#1{#1}\fi
\expandafter\ifx\csname url\endcsname\relax
  \def\url#1{\texttt{#1}}\fi
\expandafter\ifx\csname urlprefix\endcsname\relax\def\urlprefix{URL }\fi
\providecommand{\bibinfo}[2]{#2}
\providecommand{\eprint}[2][]{\url{#2}}

\bibitem[{\citenamefont{Shockley and Queisser}(1961)}]{10.1063/1.1736034}
\bibinfo{author}{\bibfnamefont{W.}~\bibnamefont{Shockley}} \bibnamefont{and}
  \bibinfo{author}{\bibfnamefont{H.}~\bibnamefont{Queisser}},
  \bibinfo{journal}{J. Appl. Phys.} \textbf{\bibinfo{volume}{32}},
  \bibinfo{pages}{510} (\bibinfo{year}{1961}).

\bibitem[{\citenamefont{Ellingson
  et~al.}(2005{\natexlab{a}})\citenamefont{Ellingson, Beard, Johnson, Yu,
  Micic, Nozik, Shabaev, and Efros}}]{ISI:000229120900009}
\bibinfo{author}{\bibfnamefont{R.~J.} \bibnamefont{Ellingson}},
  \bibinfo{author}{\bibfnamefont{M.~C.} \bibnamefont{Beard}},
  \bibinfo{author}{\bibfnamefont{J.~C.} \bibnamefont{Johnson}},
  \bibinfo{author}{\bibfnamefont{P.~R.} \bibnamefont{Yu}},
  \bibinfo{author}{\bibfnamefont{O.~I.} \bibnamefont{Micic}},
  \bibinfo{author}{\bibfnamefont{A.~J.} \bibnamefont{Nozik}},
  \bibinfo{author}{\bibfnamefont{A.}~\bibnamefont{Shabaev}}, \bibnamefont{and}
  \bibinfo{author}{\bibfnamefont{A.~L.} \bibnamefont{Efros}},
  \bibinfo{journal}{Nano Letters} \textbf{\bibinfo{volume}{5}},
  \bibinfo{pages}{865} (\bibinfo{year}{2005}{\natexlab{a}}).

\bibitem[{\citenamefont{Nozik}(2002)}]{AJ2002115}
\bibinfo{author}{\bibfnamefont{A.~J.} \bibnamefont{Nozik}},
  \bibinfo{journal}{Physica E: Low-dimensional Systems and Nanostructures}
  \textbf{\bibinfo{volume}{14}}, \bibinfo{pages}{115 } (\bibinfo{year}{2002}).

\bibitem[{\citenamefont{de~Boer et~al.}(2013)\citenamefont{de~Boer, de~Jong,
  Timmerman, Gregorkiewicz, Zhang, Buma, Poddubny, Prokofiev, and
  Yassievich}}]{PhysRevB.88.155304}
\bibinfo{author}{\bibfnamefont{W.~D. A.~M.} \bibnamefont{de~Boer}},
  \bibinfo{author}{\bibfnamefont{E.~M. L.~D.} \bibnamefont{de~Jong}},
  \bibinfo{author}{\bibfnamefont{D.}~\bibnamefont{Timmerman}},
  \bibinfo{author}{\bibfnamefont{T.}~\bibnamefont{Gregorkiewicz}},
  \bibinfo{author}{\bibfnamefont{H.}~\bibnamefont{Zhang}},
  \bibinfo{author}{\bibfnamefont{W.~J.} \bibnamefont{Buma}},
  \bibinfo{author}{\bibfnamefont{A.~N.} \bibnamefont{Poddubny}},
  \bibinfo{author}{\bibfnamefont{A.~A.} \bibnamefont{Prokofiev}},
  \bibnamefont{and} \bibinfo{author}{\bibfnamefont{I.~N.}
  \bibnamefont{Yassievich}}, \bibinfo{journal}{Phys. Rev. B}
  \textbf{\bibinfo{volume}{88}}, \bibinfo{pages}{155304}
  (\bibinfo{year}{2013}).

\bibitem[{\citenamefont{Stewart et~al.}(2013)\citenamefont{Stewart, Padilha,
  Bae, Koh, Pietryga, and Klimov}}]{doi:10.1021/jz4004334}
\bibinfo{author}{\bibfnamefont{J.}~\bibnamefont{Stewart}},
  \bibinfo{author}{\bibfnamefont{L.}~\bibnamefont{Padilha}},
  \bibinfo{author}{\bibfnamefont{W.}~\bibnamefont{Bae}},
  \bibinfo{author}{\bibfnamefont{W.}~\bibnamefont{Koh}},
  \bibinfo{author}{\bibfnamefont{J.}~\bibnamefont{Pietryga}}, \bibnamefont{and}
  \bibinfo{author}{\bibfnamefont{V.}~\bibnamefont{Klimov}},
  \bibinfo{journal}{The Journal of Physical Chemistry Letters}
  \textbf{\bibinfo{volume}{4}}, \bibinfo{pages}{2061} (\bibinfo{year}{2013}).

\bibitem[{\citenamefont{Bude and Hess}(1992)}]{5144200}
\bibinfo{author}{\bibfnamefont{J.}~\bibnamefont{Bude}} \bibnamefont{and}
  \bibinfo{author}{\bibfnamefont{K.}~\bibnamefont{Hess}},
  \bibinfo{journal}{Journal of Applied Physics} \textbf{\bibinfo{volume}{72}},
  \bibinfo{pages}{3554 } (\bibinfo{year}{1992}).

\bibitem[{\citenamefont{Jung et~al.}(1996)\citenamefont{Jung, Taniguchi, and
  Hamaguchi}}]{5014421}
\bibinfo{author}{\bibfnamefont{H.~K.} \bibnamefont{Jung}},
  \bibinfo{author}{\bibfnamefont{K.}~\bibnamefont{Taniguchi}},
  \bibnamefont{and}
  \bibinfo{author}{\bibfnamefont{C.}~\bibnamefont{Hamaguchi}},
  \bibinfo{journal}{Journal of Applied Physics}  (\bibinfo{year}{1996}).

\bibitem[{\citenamefont{Harrison et~al.}(1999)\citenamefont{Harrison, Abram,
  and Brand}}]{10.1063/1.370658}
\bibinfo{author}{\bibfnamefont{D.}~\bibnamefont{Harrison}},
  \bibinfo{author}{\bibfnamefont{R.~A.} \bibnamefont{Abram}}, \bibnamefont{and}
  \bibinfo{author}{\bibfnamefont{S.}~\bibnamefont{Brand}},
  \bibinfo{journal}{AIP} \textbf{\bibinfo{volume}{85}}, \bibinfo{pages}{8186}
  (\bibinfo{year}{1999}).

\bibitem[{\citenamefont{Nozik}(2001)}]{doi:10.1146/annurev.physchem.52.1.193}
\bibinfo{author}{\bibfnamefont{A.}~\bibnamefont{Nozik}},
  \bibinfo{journal}{Annual Review of Physical Chemistry}
  \textbf{\bibinfo{volume}{52}}, \bibinfo{pages}{193} (\bibinfo{year}{2001}).

\bibitem[{\citenamefont{Ellingson
  et~al.}(2005{\natexlab{b}})\citenamefont{Ellingson, Beard, Johnson, Yu,
  Micic, Nozik, Shabaev, and Efros}}]{doi:10.1021/nl0502672}
\bibinfo{author}{\bibfnamefont{R.}~\bibnamefont{Ellingson}},
  \bibinfo{author}{\bibfnamefont{M.}~\bibnamefont{Beard}},
  \bibinfo{author}{\bibfnamefont{J.}~\bibnamefont{Johnson}},
  \bibinfo{author}{\bibfnamefont{P.}~\bibnamefont{Yu}},
  \bibinfo{author}{\bibfnamefont{O.}~\bibnamefont{Micic}},
  \bibinfo{author}{\bibfnamefont{A.}~\bibnamefont{Nozik}},
  \bibinfo{author}{\bibfnamefont{A.}~\bibnamefont{Shabaev}}, \bibnamefont{and}
  \bibinfo{author}{\bibfnamefont{A.}~\bibnamefont{Efros}},
  \bibinfo{journal}{Nano Letters} \textbf{\bibinfo{volume}{5}},
  \bibinfo{pages}{865} (\bibinfo{year}{2005}{\natexlab{b}}).

\bibitem[{\citenamefont{McGuire et~al.}(2010)\citenamefont{McGuire, Sykora,
  Joo, Pietryga, and Klimov}}]{doi:10.1021/nl100177c}
\bibinfo{author}{\bibfnamefont{J.}~\bibnamefont{McGuire}},
  \bibinfo{author}{\bibfnamefont{M.}~\bibnamefont{Sykora}},
  \bibinfo{author}{\bibfnamefont{J.}~\bibnamefont{Joo}},
  \bibinfo{author}{\bibfnamefont{J.}~\bibnamefont{Pietryga}}, \bibnamefont{and}
  \bibinfo{author}{\bibfnamefont{V.}~\bibnamefont{Klimov}},
  \bibinfo{journal}{Nano Letters} \textbf{\bibinfo{volume}{10}},
  \bibinfo{pages}{2049} (\bibinfo{year}{2010}).

\bibitem[{\citenamefont{Gabor}(2013)}]{doi:10.1021/ar300189j}
\bibinfo{author}{\bibfnamefont{N.~M.} \bibnamefont{Gabor}},
  \bibinfo{journal}{Accounts of Chemical Research}
  \textbf{\bibinfo{volume}{46}}, \bibinfo{pages}{1348} (\bibinfo{year}{2013}).

\bibitem[{\citenamefont{Semonin et~al.}(2011)\citenamefont{Semonin, Luther,
  Choi, Chen, Gao, Nozik, and Beard}}]{Semonin16122011}
\bibinfo{author}{\bibfnamefont{O.}~\bibnamefont{Semonin}},
  \bibinfo{author}{\bibfnamefont{J.}~\bibnamefont{Luther}},
  \bibinfo{author}{\bibfnamefont{S.}~\bibnamefont{Choi}},
  \bibinfo{author}{\bibfnamefont{H.-Y.} \bibnamefont{Chen}},
  \bibinfo{author}{\bibfnamefont{J.}~\bibnamefont{Gao}},
  \bibinfo{author}{\bibfnamefont{A.~J.} \bibnamefont{Nozik}}, \bibnamefont{and}
  \bibinfo{author}{\bibfnamefont{M.~C.} \bibnamefont{Beard}},
  \bibinfo{journal}{Science} \textbf{\bibinfo{volume}{334}},
  \bibinfo{pages}{1530} (\bibinfo{year}{2011}).

\bibitem[{\citenamefont{Wang et~al.}(2010)\citenamefont{Wang, Khafizov, Tu,
  Zheng, and Krauss}}]{doi:10.1021/nl100343j}
\bibinfo{author}{\bibfnamefont{S.}~\bibnamefont{Wang}},
  \bibinfo{author}{\bibfnamefont{M.}~\bibnamefont{Khafizov}},
  \bibinfo{author}{\bibfnamefont{X.}~\bibnamefont{Tu}},
  \bibinfo{author}{\bibfnamefont{M.}~\bibnamefont{Zheng}}, \bibnamefont{and}
  \bibinfo{author}{\bibfnamefont{T.}~\bibnamefont{Krauss}},
  \bibinfo{journal}{Nano Letters} \textbf{\bibinfo{volume}{10}},
  \bibinfo{pages}{2381} (\bibinfo{year}{2010}).

\bibitem[{\citenamefont{Gabor et~al.}(2009)\citenamefont{Gabor, Zhong, Bosnick,
  Park, and McEuen}}]{Gabor11092009}
\bibinfo{author}{\bibfnamefont{N.}~\bibnamefont{Gabor}},
  \bibinfo{author}{\bibfnamefont{Z.}~\bibnamefont{Zhong}},
  \bibinfo{author}{\bibfnamefont{K.}~\bibnamefont{Bosnick}},
  \bibinfo{author}{\bibfnamefont{J.}~\bibnamefont{Park}}, \bibnamefont{and}
  \bibinfo{author}{\bibfnamefont{P.}~\bibnamefont{McEuen}},
  \bibinfo{journal}{Science} \textbf{\bibinfo{volume}{325}},
  \bibinfo{pages}{1367} (\bibinfo{year}{2009}).

\bibitem[{\citenamefont{Perebeinos and Avouris}(2006)}]{PhysRevB.74.121410}
\bibinfo{author}{\bibfnamefont{V.}~\bibnamefont{Perebeinos}} \bibnamefont{and}
  \bibinfo{author}{\bibfnamefont{P.}~\bibnamefont{Avouris}},
  \bibinfo{journal}{Phys. Rev. B} \textbf{\bibinfo{volume}{74}},
  \bibinfo{pages}{121410} (\bibinfo{year}{2006}).

\bibitem[{\citenamefont{Konabe and Okada}(2012)}]{PhysRevLett.108.227401}
\bibinfo{author}{\bibfnamefont{S.}~\bibnamefont{Konabe}} \bibnamefont{and}
  \bibinfo{author}{\bibfnamefont{S.}~\bibnamefont{Okada}},
  \bibinfo{journal}{Phys. Rev. Lett.} \textbf{\bibinfo{volume}{108}},
  \bibinfo{pages}{227401} (\bibinfo{year}{2012}).

\bibitem[{\citenamefont{Velizhanin and
  Piryatinski}(2011)}]{PhysRevLett.106.207401}
\bibinfo{author}{\bibfnamefont{K.}~\bibnamefont{Velizhanin}} \bibnamefont{and}
  \bibinfo{author}{\bibfnamefont{A.}~\bibnamefont{Piryatinski}},
  \bibinfo{journal}{Phys. Rev. Lett.} \textbf{\bibinfo{volume}{106}},
  \bibinfo{pages}{207401} (\bibinfo{year}{2011}).

\bibitem[{\citenamefont{Velizhanin and Piryatinski}(2012)}]{PhysRevB.86.165319}
\bibinfo{author}{\bibfnamefont{K.~A.} \bibnamefont{Velizhanin}}
  \bibnamefont{and}
  \bibinfo{author}{\bibfnamefont{A.}~\bibnamefont{Piryatinski}},
  \bibinfo{journal}{Phys. Rev. B} \textbf{\bibinfo{volume}{86}},
  \bibinfo{pages}{165319} (\bibinfo{year}{2012}).

\bibitem[{\citenamefont{Kilina et~al.}(2015)\citenamefont{Kilina, Kilin, and
  Tretiak}}]{doi:10.1021/acs.chemrev.5b00012}
\bibinfo{author}{\bibfnamefont{S.}~\bibnamefont{Kilina}},
  \bibinfo{author}{\bibfnamefont{D.}~\bibnamefont{Kilin}}, \bibnamefont{and}
  \bibinfo{author}{\bibfnamefont{S.}~\bibnamefont{Tretiak}},
  \bibinfo{journal}{Chemical Reviews} \textbf{\bibinfo{volume}{115}},
  \bibinfo{pages}{5929} (\bibinfo{year}{2015}).

\bibitem[{\citenamefont{Kryjevski et~al.}(2016)\citenamefont{Kryjevski,
  Gifford, Kilina, and Kilin}}]{doi:10.1063/1.4963735}
\bibinfo{author}{\bibfnamefont{A.}~\bibnamefont{Kryjevski}},
  \bibinfo{author}{\bibfnamefont{B.}~\bibnamefont{Gifford}},
  \bibinfo{author}{\bibfnamefont{S.}~\bibnamefont{Kilina}}, \bibnamefont{and}
  \bibinfo{author}{\bibfnamefont{D.}~\bibnamefont{Kilin}},
  \bibinfo{journal}{The Journal of Chemical Physics}
  \textbf{\bibinfo{volume}{145}}, \bibinfo{pages}{154112}
  (\bibinfo{year}{2016}).

\bibitem[{\citenamefont{Smith and Michl}(2010)}]{doi:10.1021/cr1002613}
\bibinfo{author}{\bibfnamefont{M.~B.} \bibnamefont{Smith}} \bibnamefont{and}
  \bibinfo{author}{\bibfnamefont{J.}~\bibnamefont{Michl}},
  \bibinfo{journal}{Chemical Reviews} \textbf{\bibinfo{volume}{110}},
  \bibinfo{pages}{6891} (\bibinfo{year}{2010}).

\bibitem[{\citenamefont{Lee et~al.}(2013)\citenamefont{Lee, Jadhav, Reusswig,
  Yost, Thompson, Congreve, Hontz, Van~Voorhis, and
  Baldo}}]{doi:10.1021/ar300288e}
\bibinfo{author}{\bibfnamefont{J.}~\bibnamefont{Lee}},
  \bibinfo{author}{\bibfnamefont{P.}~\bibnamefont{Jadhav}},
  \bibinfo{author}{\bibfnamefont{P.~D.} \bibnamefont{Reusswig}},
  \bibinfo{author}{\bibfnamefont{S.~R.} \bibnamefont{Yost}},
  \bibinfo{author}{\bibfnamefont{N.~J.} \bibnamefont{Thompson}},
  \bibinfo{author}{\bibfnamefont{D.~N.} \bibnamefont{Congreve}},
  \bibinfo{author}{\bibfnamefont{E.}~\bibnamefont{Hontz}},
  \bibinfo{author}{\bibfnamefont{T.}~\bibnamefont{Van~Voorhis}},
  \bibnamefont{and} \bibinfo{author}{\bibfnamefont{M.~A.} \bibnamefont{Baldo}},
  \bibinfo{journal}{Accounts of Chemical Research}
  \textbf{\bibinfo{volume}{46}}, \bibinfo{pages}{1300} (\bibinfo{year}{2013}).

\bibitem[{\citenamefont{Tretiak}(2007)}]{doi:10.1021/nl070355h}
\bibinfo{author}{\bibfnamefont{S.}~\bibnamefont{Tretiak}},
  \bibinfo{journal}{Nano Letters} \textbf{\bibinfo{volume}{7}},
  \bibinfo{pages}{2201} (\bibinfo{year}{2007}).

\bibitem[{\citenamefont{Wang et~al.}(2016)\citenamefont{Wang, Garcia, Monaco,
  Schatschneider, and Marom}}]{C6CE00873A}
\bibinfo{author}{\bibfnamefont{X.}~\bibnamefont{Wang}},
  \bibinfo{author}{\bibfnamefont{T.}~\bibnamefont{Garcia}},
  \bibinfo{author}{\bibfnamefont{S.}~\bibnamefont{Monaco}},
  \bibinfo{author}{\bibfnamefont{B.}~\bibnamefont{Schatschneider}},
  \bibnamefont{and} \bibinfo{author}{\bibfnamefont{N.}~\bibnamefont{Marom}},
  \bibinfo{journal}{CrystEngComm} \textbf{\bibinfo{volume}{18}},
  \bibinfo{pages}{7353} (\bibinfo{year}{2016}).

\bibitem[{\citenamefont{Stich et~al.}(2014)\citenamefont{Stich, Spath, Kraus,
  Sperlich, Dyakonov, and Hertel}}]{nat_phot_2014_s1cnt}
\bibinfo{author}{\bibfnamefont{D.}~\bibnamefont{Stich}},
  \bibinfo{author}{\bibfnamefont{F.}~\bibnamefont{Spath}},
  \bibinfo{author}{\bibfnamefont{H.}~\bibnamefont{Kraus}},
  \bibinfo{author}{\bibfnamefont{A.}~\bibnamefont{Sperlich}},
  \bibinfo{author}{\bibfnamefont{V.}~\bibnamefont{Dyakonov}}, \bibnamefont{and}
  \bibinfo{author}{\bibfnamefont{T.}~\bibnamefont{Hertel}},
  \bibinfo{journal}{Nature Photonics} \textbf{\bibinfo{volume}{8}},
  \bibinfo{pages}{139} (\bibinfo{year}{2014}).

\bibitem[{\citenamefont{Fetter and Walecka}(1971)}]{FW}
\bibinfo{author}{\bibfnamefont{A.~L.} \bibnamefont{Fetter}} \bibnamefont{and}
  \bibinfo{author}{\bibfnamefont{J.}~\bibnamefont{Walecka}},
  \emph{\bibinfo{title}{Quantum Theory of Many-Particle Systems}}
  (\bibinfo{publisher}{McGraw-Hill}, \bibinfo{address}{New York},
  \bibinfo{year}{1971}).

\bibitem[{\citenamefont{Mahan}(1993)}]{Mahan}
\bibinfo{author}{\bibfnamefont{G.}~\bibnamefont{Mahan}},
  \emph{\bibinfo{title}{Many-Particle Physics}} (\bibinfo{publisher}{Plenum},
  \bibinfo{address}{New York, N.Y.}, \bibinfo{year}{1993}),
  \bibinfo{edition}{2nd} ed.

\bibitem[{\citenamefont{Kryjevski and Kilin}(2014)}]{molphysKK}
\bibinfo{author}{\bibfnamefont{A.}~\bibnamefont{Kryjevski}} \bibnamefont{and}
  \bibinfo{author}{\bibfnamefont{D.}~\bibnamefont{Kilin}},
  \bibinfo{journal}{Molecular Physics} \textbf{\bibinfo{volume}{112}},
  \bibinfo{pages}{430} (\bibinfo{year}{2014}).

\bibitem[{\citenamefont{Onida et~al.}(2002)\citenamefont{Onida, Reining, and
  Rubio}}]{RevModPhys.74.601}
\bibinfo{author}{\bibfnamefont{G.}~\bibnamefont{Onida}},
  \bibinfo{author}{\bibfnamefont{L.}~\bibnamefont{Reining}}, \bibnamefont{and}
  \bibinfo{author}{\bibfnamefont{A.}~\bibnamefont{Rubio}},
  \bibinfo{journal}{Rev. Mod. Phys.} \textbf{\bibinfo{volume}{74}},
  \bibinfo{pages}{601} (\bibinfo{year}{2002}).

\bibitem[{\citenamefont{K\"ummel and Kronik}(2008)}]{RevModPhys.80.3}
\bibinfo{author}{\bibfnamefont{S.}~\bibnamefont{K\"ummel}} \bibnamefont{and}
  \bibinfo{author}{\bibfnamefont{L.}~\bibnamefont{Kronik}},
  \bibinfo{journal}{Rev. Mod. Phys.} \textbf{\bibinfo{volume}{80}},
  \bibinfo{pages}{3} (\bibinfo{year}{2008}).

\bibitem[{\citenamefont{Rohlfing and Louie}(2000)}]{PhysRevB.62.4927}
\bibinfo{author}{\bibfnamefont{M.}~\bibnamefont{Rohlfing}} \bibnamefont{and}
  \bibinfo{author}{\bibfnamefont{S.}~\bibnamefont{Louie}},
  \bibinfo{journal}{Phys. Rev. B} \textbf{\bibinfo{volume}{62}},
  \bibinfo{pages}{4927} (\bibinfo{year}{2000}).

\bibitem[{\citenamefont{Strinati}(1984)}]{PhysRevB.29.5718}
\bibinfo{author}{\bibfnamefont{G.}~\bibnamefont{Strinati}},
  \bibinfo{journal}{Phys. Rev. B} \textbf{\bibinfo{volume}{29}},
  \bibinfo{pages}{5718} (\bibinfo{year}{1984}).

\bibitem[{\citenamefont{Berestetskii et~al.}(1979)\citenamefont{Berestetskii,
  Lifshitz, and Pitaevskii}}]{Berestetskii:1979aa}
\bibinfo{author}{\bibfnamefont{V.}~\bibnamefont{Berestetskii}},
  \bibinfo{author}{\bibfnamefont{E.}~\bibnamefont{Lifshitz}}, \bibnamefont{and}
  \bibinfo{author}{\bibfnamefont{L.}~\bibnamefont{Pitaevskii}},
  \emph{\bibinfo{title}{Quantum Electrodynamics}} (\bibinfo{publisher}{Oxford,
  U.K.: Pergamon Press}, \bibinfo{year}{1979}).

\bibitem[{\citenamefont{Beane et~al.}(2000)\citenamefont{Beane, Bedaque,
  Haxton, Phillips, and Savage}}]{Beane:2000fx}
\bibinfo{author}{\bibfnamefont{S.}~\bibnamefont{Beane}},
  \bibinfo{author}{\bibfnamefont{P.}~\bibnamefont{Bedaque}},
  \bibinfo{author}{\bibfnamefont{W.}~\bibnamefont{Haxton}},
  \bibinfo{author}{\bibfnamefont{D.}~\bibnamefont{Phillips}}, \bibnamefont{and}
  \bibinfo{author}{\bibfnamefont{M.}~\bibnamefont{Savage}},
  \bibinfo{journal}{Shifman, M. (ed.): At the frontier of particle physics}
  \textbf{\bibinfo{volume}{1}}, \bibinfo{pages}{133} (\bibinfo{year}{2000}).

\bibitem[{\citenamefont{Spataru et~al.}(2004)\citenamefont{Spataru,
  Ismail-Beigi, Benedict, and Louie}}]{PhysRevLett.92.077402}
\bibinfo{author}{\bibfnamefont{C.}~\bibnamefont{Spataru}},
  \bibinfo{author}{\bibfnamefont{S.}~\bibnamefont{Ismail-Beigi}},
  \bibinfo{author}{\bibfnamefont{L.}~\bibnamefont{Benedict}}, \bibnamefont{and}
  \bibinfo{author}{\bibfnamefont{S.}~\bibnamefont{Louie}},
  \bibinfo{journal}{Phys. Rev. Lett.} \textbf{\bibinfo{volume}{92}},
  \bibinfo{pages}{077402} (\bibinfo{year}{2004}).

\bibitem[{\citenamefont{Perebeinos et~al.}(2004)\citenamefont{Perebeinos,
  Tersoff, and Avouris}}]{PhysRevLett.92.257402}
\bibinfo{author}{\bibfnamefont{V.}~\bibnamefont{Perebeinos}},
  \bibinfo{author}{\bibfnamefont{J.}~\bibnamefont{Tersoff}}, \bibnamefont{and}
  \bibinfo{author}{\bibfnamefont{P.}~\bibnamefont{Avouris}},
  \bibinfo{journal}{Phys. Rev. Lett.} \textbf{\bibinfo{volume}{92}},
  \bibinfo{pages}{257402} (\bibinfo{year}{2004}).

\bibitem[{\citenamefont{Spataru et~al.}(2005)\citenamefont{Spataru,
  Ismail-Beigi, Capaz, and Louie}}]{PhysRevLett.95.247402}
\bibinfo{author}{\bibfnamefont{C.}~\bibnamefont{Spataru}},
  \bibinfo{author}{\bibfnamefont{S.}~\bibnamefont{Ismail-Beigi}},
  \bibinfo{author}{\bibfnamefont{R.}~\bibnamefont{Capaz}}, \bibnamefont{and}
  \bibinfo{author}{\bibfnamefont{S.}~\bibnamefont{Louie}},
  \bibinfo{journal}{Phys. Rev. Lett.} \textbf{\bibinfo{volume}{95}},
  \bibinfo{pages}{247402} (\bibinfo{year}{2005}).

\bibitem[{\citenamefont{\"O\ifmmode~\breve{g}\else \u{g}\fi{}\"ut
  et~al.}(2003)\citenamefont{\"O\ifmmode~\breve{g}\else \u{g}\fi{}\"ut,
  Burdick, Saad, and Chelikowsky}}]{PhysRevLett.90.127401}
\bibinfo{author}{\bibfnamefont{S.}~\bibnamefont{\"O\ifmmode~\breve{g}\else
  \u{g}\fi{}\"ut}}, \bibinfo{author}{\bibfnamefont{R.}~\bibnamefont{Burdick}},
  \bibinfo{author}{\bibfnamefont{Y.}~\bibnamefont{Saad}}, \bibnamefont{and}
  \bibinfo{author}{\bibfnamefont{J.}~\bibnamefont{Chelikowsky}},
  \bibinfo{journal}{Phys. Rev. Lett.} \textbf{\bibinfo{volume}{90}},
  \bibinfo{pages}{127401} (\bibinfo{year}{2003}).

\bibitem[{\citenamefont{Benedict et~al.}(2003)\citenamefont{Benedict, Puzder,
  Williamson, Grossman, Galli, Klepeis, Raty, and
  Pankratov}}]{PhysRevB.68.085310}
\bibinfo{author}{\bibfnamefont{L.}~\bibnamefont{Benedict}},
  \bibinfo{author}{\bibfnamefont{A.}~\bibnamefont{Puzder}},
  \bibinfo{author}{\bibfnamefont{A.}~\bibnamefont{Williamson}},
  \bibinfo{author}{\bibfnamefont{J.}~\bibnamefont{Grossman}},
  \bibinfo{author}{\bibfnamefont{G.}~\bibnamefont{Galli}},
  \bibinfo{author}{\bibfnamefont{J.}~\bibnamefont{Klepeis}},
  \bibinfo{author}{\bibfnamefont{J.-Y.} \bibnamefont{Raty}}, \bibnamefont{and}
  \bibinfo{author}{\bibfnamefont{O.}~\bibnamefont{Pankratov}},
  \bibinfo{journal}{Phys. Rev. B} \textbf{\bibinfo{volume}{68}},
  \bibinfo{pages}{085310} (\bibinfo{year}{2003}).

\bibitem[{\citenamefont{Wilson et~al.}(2009)\citenamefont{Wilson, Lu, Gygi, and
  Galli}}]{PhysRevB.79.245106}
\bibinfo{author}{\bibfnamefont{H.}~\bibnamefont{Wilson}},
  \bibinfo{author}{\bibfnamefont{D.}~\bibnamefont{Lu}},
  \bibinfo{author}{\bibfnamefont{F.}~\bibnamefont{Gygi}}, \bibnamefont{and}
  \bibinfo{author}{\bibfnamefont{G.}~\bibnamefont{Galli}},
  \bibinfo{journal}{Phys. Rev. B} \textbf{\bibinfo{volume}{79}},
  \bibinfo{pages}{245106} (\bibinfo{year}{2009}).

\bibitem[{\citenamefont{Rohlfing and Louie}(1998)}]{PhysRevLett.80.3320}
\bibinfo{author}{\bibfnamefont{M.}~\bibnamefont{Rohlfing}} \bibnamefont{and}
  \bibinfo{author}{\bibfnamefont{S.~G.} \bibnamefont{Louie}},
  \bibinfo{journal}{Phys. Rev. Lett.} \textbf{\bibinfo{volume}{80}},
  \bibinfo{pages}{3320} (\bibinfo{year}{1998}).

\bibitem[{\citenamefont{Vydrov et~al.}(2006)\citenamefont{Vydrov, Heyd, Krukau,
  and Scuseria}}]{vydrov:074106}
\bibinfo{author}{\bibfnamefont{O.}~\bibnamefont{Vydrov}},
  \bibinfo{author}{\bibfnamefont{J.}~\bibnamefont{Heyd}},
  \bibinfo{author}{\bibfnamefont{A.}~\bibnamefont{Krukau}}, \bibnamefont{and}
  \bibinfo{author}{\bibfnamefont{G.}~\bibnamefont{Scuseria}},
  \bibinfo{journal}{The Journal of Chemical Physics}
  \textbf{\bibinfo{volume}{125}}, \bibinfo{eid}{074106} (\bibinfo{year}{2006}).

\bibitem[{\citenamefont{Heyd et~al.}(2006)\citenamefont{Heyd, Scuseria, and
  Ernzerhof}}]{heyd:219906}
\bibinfo{author}{\bibfnamefont{J.}~\bibnamefont{Heyd}},
  \bibinfo{author}{\bibfnamefont{G.}~\bibnamefont{Scuseria}}, \bibnamefont{and}
  \bibinfo{author}{\bibfnamefont{M.}~\bibnamefont{Ernzerhof}},
  \bibinfo{journal}{The Journal of Chemical Physics}
  \textbf{\bibinfo{volume}{124}}, \bibinfo{eid}{219906} (\bibinfo{year}{2006}).

\bibitem[{\citenamefont{Govoni and Galli}(2015)}]{doi:10.1021/ct500958p}
\bibinfo{author}{\bibfnamefont{M.}~\bibnamefont{Govoni}} \bibnamefont{and}
  \bibinfo{author}{\bibfnamefont{G.}~\bibnamefont{Galli}},
  \bibinfo{journal}{Journal of Chemical Theory and Computation}
  \textbf{\bibinfo{volume}{11}}, \bibinfo{pages}{2680} (\bibinfo{year}{2015}).

\bibitem[{\citenamefont{Jain et~al.}(2011)\citenamefont{Jain, Chelikowsky, and
  Louie}}]{PhysRevLett.107.216806}
\bibinfo{author}{\bibfnamefont{M.}~\bibnamefont{Jain}},
  \bibinfo{author}{\bibfnamefont{J.~R.} \bibnamefont{Chelikowsky}},
  \bibnamefont{and} \bibinfo{author}{\bibfnamefont{S.~G.} \bibnamefont{Louie}},
  \bibinfo{journal}{Phys. Rev. Lett.} \textbf{\bibinfo{volume}{107}},
  \bibinfo{pages}{216806} (\bibinfo{year}{2011}).

\bibitem[{\citenamefont{Hybertsen and Louie}(1986)}]{PhysRevB.34.5390}
\bibinfo{author}{\bibfnamefont{M.}~\bibnamefont{Hybertsen}} \bibnamefont{and}
  \bibinfo{author}{\bibfnamefont{S.}~\bibnamefont{Louie}},
  \bibinfo{journal}{Phys. Rev. B} \textbf{\bibinfo{volume}{34}},
  \bibinfo{pages}{5390} (\bibinfo{year}{1986}).

\bibitem[{\citenamefont{Abrikosov et~al.}(1963)\citenamefont{Abrikosov, Gorkov,
  and Dzyaloshinski}}]{AGD}
\bibinfo{author}{\bibfnamefont{A.~A.} \bibnamefont{Abrikosov}},
  \bibinfo{author}{\bibfnamefont{L.}~\bibnamefont{Gorkov}}, \bibnamefont{and}
  \bibinfo{author}{\bibfnamefont{I.~E.} \bibnamefont{Dzyaloshinski}},
  \emph{\bibinfo{title}{Methods of Quantum Field Theory in Statistical
  Physics}} (\bibinfo{publisher}{Prentice-Hall}, \bibinfo{address}{Englewood
  Cliffs, NJ}, \bibinfo{year}{1963}).

\bibitem[{\citenamefont{Kryjevski and
  Kilin}(2016)}]{doi:10.1080/00268976.2015.1076580}
\bibinfo{author}{\bibfnamefont{A.}~\bibnamefont{Kryjevski}} \bibnamefont{and}
  \bibinfo{author}{\bibfnamefont{D.}~\bibnamefont{Kilin}},
  \bibinfo{journal}{Molecular Physics} \textbf{\bibinfo{volume}{114}},
  \bibinfo{pages}{365} (\bibinfo{year}{2016}).

\bibitem[{\citenamefont{Deslippe et~al.}(2012)\citenamefont{Deslippe,
  Samsonidze, Strubbe, Jain, Cohen, and Louie}}]{Deslippe20121269}
\bibinfo{author}{\bibfnamefont{J.}~\bibnamefont{Deslippe}},
  \bibinfo{author}{\bibfnamefont{G.}~\bibnamefont{Samsonidze}},
  \bibinfo{author}{\bibfnamefont{D.}~\bibnamefont{Strubbe}},
  \bibinfo{author}{\bibfnamefont{M.}~\bibnamefont{Jain}},
  \bibinfo{author}{\bibfnamefont{M.}~\bibnamefont{Cohen}}, \bibnamefont{and}
  \bibinfo{author}{\bibfnamefont{S.}~\bibnamefont{Louie}},
  \bibinfo{journal}{Computer Physics Communications}
  \textbf{\bibinfo{volume}{183}}, \bibinfo{pages}{1269 }
  (\bibinfo{year}{2012}).

\bibitem[{\citenamefont{Weisman and Bachilo}(2003)}]{doi:10.1021/nl034428i}
\bibinfo{author}{\bibfnamefont{R.~B.} \bibnamefont{Weisman}} \bibnamefont{and}
  \bibinfo{author}{\bibfnamefont{S.}~\bibnamefont{Bachilo}},
  \bibinfo{journal}{Nano Letters} \textbf{\bibinfo{volume}{3}},
  \bibinfo{pages}{1235} (\bibinfo{year}{2003}).

\bibitem[{\citenamefont{Berkelbach et~al.}(2013)\citenamefont{Berkelbach,
  Hybertsen, and Reichman}}]{doi:10.1063/1.4794425}
\bibinfo{author}{\bibfnamefont{T.~C.} \bibnamefont{Berkelbach}},
  \bibinfo{author}{\bibfnamefont{M.~S.} \bibnamefont{Hybertsen}},
  \bibnamefont{and} \bibinfo{author}{\bibfnamefont{D.~R.}
  \bibnamefont{Reichman}}, \bibinfo{journal}{The Journal of Chemical Physics}
  \textbf{\bibinfo{volume}{138}}, \bibinfo{pages}{114102}
  (\bibinfo{year}{2013}).

\bibitem[{\citenamefont{Bl\"ochl}(1994)}]{PhysRevB.50.17953}
\bibinfo{author}{\bibfnamefont{P.~E.} \bibnamefont{Bl\"ochl}},
  \bibinfo{journal}{Phys. Rev. B} \textbf{\bibinfo{volume}{50}},
  \bibinfo{pages}{17953} (\bibinfo{year}{1994}).

\bibitem[{\citenamefont{Kresse and Joubert}(1999)}]{PhysRevB.59.1758}
\bibinfo{author}{\bibfnamefont{G.}~\bibnamefont{Kresse}} \bibnamefont{and}
  \bibinfo{author}{\bibfnamefont{D.}~\bibnamefont{Joubert}},
  \bibinfo{journal}{Phys. Rev. B} \textbf{\bibinfo{volume}{59}},
  \bibinfo{pages}{1758} (\bibinfo{year}{1999}).

\bibitem[{\citenamefont{Lifshitz and Pitaevskii}(1981)}]{Landau10}
\bibinfo{author}{\bibfnamefont{E.~M.} \bibnamefont{Lifshitz}} \bibnamefont{and}
  \bibinfo{author}{\bibfnamefont{L.~P.} \bibnamefont{Pitaevskii}},
  \emph{\bibinfo{title}{Physical Kinetics}} (\bibinfo{publisher}{Pergamon
  Press}, \bibinfo{address}{New York}, \bibinfo{year}{1981}),
  \bibinfo{edition}{1st} ed.

\bibitem[{\citenamefont{Dahnovsky}(2011)}]{PhysRevB.83.165306}
\bibinfo{author}{\bibfnamefont{Y.}~\bibnamefont{Dahnovsky}},
  \bibinfo{journal}{Phys. Rev. B} \textbf{\bibinfo{volume}{83}},
  \bibinfo{pages}{165306} (\bibinfo{year}{2011}).

\bibitem[{\citenamefont{Bernardi et~al.}(2014)\citenamefont{Bernardi,
  Vigil-Fowler, Lischner, Neaton, and Louie}}]{PhysRevLett.112.257402}
\bibinfo{author}{\bibfnamefont{M.}~\bibnamefont{Bernardi}},
  \bibinfo{author}{\bibfnamefont{D.}~\bibnamefont{Vigil-Fowler}},
  \bibinfo{author}{\bibfnamefont{J.}~\bibnamefont{Lischner}},
  \bibinfo{author}{\bibfnamefont{J.~B.} \bibnamefont{Neaton}},
  \bibnamefont{and} \bibinfo{author}{\bibfnamefont{S.~G.} \bibnamefont{Louie}},
  \bibinfo{journal}{Phys. Rev. Lett.} \textbf{\bibinfo{volume}{112}},
  \bibinfo{pages}{257402} (\bibinfo{year}{2014}).

\end{thebibliography}
\end{document}